\documentclass[12pt,a4paper]{iopart}

\usepackage{iopams,subeqn}
\usepackage{graphics}
\usepackage{color}

\begin{document}

\title{Variational approximations for stochastic dynamics on graphs}

\author{A Pelizzola$^{1,2}$ and M Pretti$^{3}$\footnote{Hosted at DISAT, Politecnico di Torino, Corso Duca degli Abruzzi 24, I-10129 Torino, Italy.}}

\address{$^1$ Dipartimento di Scienza Applicata e Tecnologia (DISAT), Politecnico di Torino, Corso Duca degli Abruzzi 24, I-10129 Torino, Italy}
\address{$^2$ INFN, Sezione di Torino, Via Pietro Giuria 1, I-10125 Torino, Italy}
\address{$^3$ Consiglio Nazionale delle Ricerche - Istituto dei Sistemi Complessi (CNR-ISC)}
\ead{alessandro.pelizzola@polito.it}
\ead{marco.pretti@isc.cnr.it}

\begin{abstract}
  We investigate different mean-field-like approximations
  for stochastic dynamics on graphs, within the framework of
  a cluster-variational approach. In analogy with its equilibrium counterpart, this
  approach allows one to give a unified view of various (previously known)
  approximation schemes, and suggests quite a systematic way to improve the level of accuracy.
  We compare the different approximations with Monte Carlo simulations
  on a reversible (susceptible-infected-susceptible)
  discrete-time epidemic-spreading model on random graphs.
\end{abstract}

\maketitle

\section{Introduction}

In statistical mechanics, the \emph{cluster variational method}
(CVM), originally proposed by Kikuchi in
1951~\cite{Kikuchi1951}, is a well-established technique, which
provides a systematic way to improve the accuracy of the
simplest mean-field approximation. The classical scope of
application is that of equilibrium thermodynamics of lattice
models with discrete degrees of freedom, such as Ising or
Potts~\cite{PlischkeBergersen1994}. The basic idea is to cut
off correlations among site variables beyond the range of
certain \emph{maximal clusters} (for instance, square
``plaquettes'' on a square lattice), so that the level of
accuracy can be somehow ``tailored'' by means of a suitable
choice of such clusters. The output of the calculation is the
set of marginals of the Boltzmann distribution over the maximal
clusters. In particular, choosing single sites or
nearest-neighbor pairs as maximal clusters, the CVM recovers,
respectively, the ordinary mean-field (Bragg-Williams)
approximation and the Bethe-Peierls
approximation~\cite{PlischkeBergersen1994}. More generally, one
can show that a number of generalized mean-field
approximations, originally developed by heuristic arguments,
can be incorporated into the CVM
framework~\cite{Kikuchi1951,Pelizzola2005}. Let us also mention
the fact that the method has gained a renewed interest in
recent years, because of the formal analogy between equilibrium
statistical mechanics and a number of inference problems of
high technological relevance (error-correction decoding,
pattern recognition, etc.), which ultimately require computing
marginals of the Boltzmann distribution for (heterogeneous)
statistical-mechanical models defined on (very large)
graphs~\cite{YedidiaFreemanWeiss2005}.

In the complementary field of non-equilibrium statistical
mechanics, a rigorous foundation from first principles is still
lacking, so that a considerable part of research work is
devoted to model systems in which the (stochastic) dynamics is
assumed to be known, and properly defined to describe some kind
of phenomena. This class of models includes, for instance,
epidemic-spreading
models~\cite{WangTangStanleyBraunstein2016,PastorsatorrasCastellanoVanmieghemVespignani2015,PetermannDelosrios2004},
kinetic Ising or Ising-like
models~\cite{CrisantiSompolinsky1988} (including models of
neural-network
dynamics)~\cite{RoudiAurellHertz2009,AldanaCoppersmithKadanoff2003},
models of opinion formation and social dynamics (Voter model
and its numerous
variants)~\cite{CastellanoFortunatoLoreto2009,SchweitzerBehera2015,SucheckiEguiluzSanmiguel2005},
transport models (exclusion
processes)~\cite{ChouMallickZia2011} and so on. Increasing
research efforts have also been oriented to investigating the
aforementioned dynamics on complex networks, with special
attention to the role played by the
latter~\cite{PastorsatorrasCastellanoVanmieghemVespignani2015,BarratBarthelemyVespignani2008}.
Most of the stochastic processes involved in these models are
of the Markovian (i.e., memoryless) type, fully characterized
by a matrix of transition probabilities (stochastic matrix).
The problem can be generally rephrased in a variational form
through a \emph{thermodynamic formalism}, fully analogous to
that of equilibrium statistical mechanics, where the transition
probabilities play the role of interactions, and, within this
formalism, one can develop mean-field-like approximation
strategies along the lines of the cluster variational method.
Such a ``dynamical analog'' of the CVM was indeed put forward
in the 1960s by Kikuchi himself~\cite{Kikuchi1966}, who named
it \emph{path probability method} (PPM). The latter method has
not become so popular as its equilibrium counterpart, so that
in fact most mean-field-like approaches to stochastic dynamics
are still recently conceived on the basis of traditional
heuristic reasoning, without resorting to the
cluster-variational
machinery~\cite{WangTangStanleyBraunstein2016,PastorsatorrasCastellanoVanmieghemVespignani2015,MataFerreira2013}.

At odds with the cited works, a path-probability approach has
been recently followed by one of the
authors~\cite{Pelizzola2013}, to develop a new type of
approximation which, due to the peculiar ``shape'' of the
maximal clusters, has been called the \emph{diamond
approximation}. The latter has been applied to kinetic Ising
models (with Glauber dynamics, and either symmetric or
asymmetric couplings), but it can be straightforwardly used for
other Markov-chain models as well, provided the transition
matrix has a suitable factorized form, namely that (the
probability of) the next-time configuration of each site
variable depends only on the current-time configuration of its
neighbors. The method has been the subject of a detailed
testing work by an independent
group~\cite{DominguezDelferraroRiccitersenghi2016}, still on
various types of kinetic Ising-like models, demonstrating a
remarkable accuracy for both the transient and the steady
state, with just a slight breakdown in the notoriously
difficult case of spin glasses at low temperature. Let us
stress the fact that the derivation of the diamond
approximation in reference~\cite{Pelizzola2013} introduces a
novelty item with respect to a ``standard'' PPM
approach~\cite{WadaKaburagi1994}, in that it removes an
unnecessary constraint in the choice of maximal clusters,
potentially opening the way to new approximation schemes.

In this paper we analyze different alternative approximations
that can be worked out in the framework of a path-probability
approach along the lines of reference~\cite{Pelizzola2013}.
This analysis leads in particular to generalizing the diamond
approximation in such a way that the transition probability of
a given site variable may also depend on the current
configuration of the site itself. This step forward is
necessary for the method to be applicable to models with this
feature, such as, for instance, epidemic models or the like. We
also trace a number of connections with different approximation
schemes, previously derived either by the conventional
``heuristic'' approach or by the ``standard'' PPM approach. In
the second part of the paper, we evaluate the performance of
these different approximations in terms of both accuracy and
computational complexity, with respect to Monte Carlo
simulations. As a test model, we consider a
susceptible-infected-susceptible (SIS) epidemic model on large
random graphs, possibly with heterogeneous infection
parameters. It turns out that the generalized diamond
approximation outperforms simpler, previously known
approximation schemes, with only a slight increase in the
computational complexity.

Let us note that our investigation is carried out under the
assumption of a discrete-time dynamics, but all the
approximations can be applied as well to describe
continuous-time processes. Concerning this issue, we prove in
particular that the (generalized) diamond approximation
actually degenerates, in the continuous-time limit, into a
simpler (pairwise) approximation. From the technical point of
view, the latter result entails that, even though it is still
possible to get a more accurate approximation for
continuous-time models, one necessarily has to resort to even
larger maximal clusters, which of course requires a larger
computational effort.

The paper is organized as follows.
Section~\ref{sec:cluster-variational_approach} contains the
main theoretical issues. Sub~\ref{subsec:the_reference_model}
defines the most general model we can deal with, which we
denote as the \emph{reference model}. In
sub~\ref{subsec:thermodynamic_formalism} we introduce the
thermodynamic formalism and the basic concepts of the cluster
variational method. In sub~\ref{subsec:specific_approximations}
we describe the specific approximations which are the subject
of the present paper, in particular the one which generalizes
the diamond approximation. Most analytical details, along with
a scheme of the resulting computation procedures, are reported
in \ref{app:details} (generalized diamond approximation) and
\ref{app:other_approximations} (other approximations).
\ref{app:continuous_time} deals with the special case of
continuous-time processes, mentioned above.
Section~\ref{sec:test_models} describes the numerical tests,
whereas Section~\ref{sec:conclusions} contains a summary of the
results and some concluding remarks.

%The method has attracted new interest in more recent years,
%because of the formal analogy between equilibrium statistical
%mechanics and statistical inference. In a few words, several
%statistical inference problems of great technological relevance
%(error-correction decoding, pattern recognition, etc.)
%ultimately amount to the computation of marginals of the
%Boltzmann distribution for (heterogeneous)
%statistical-mechanical models defined on (very large) graphs.
%Very efficient (message-passing) algorithms, usually referred
%to as \emph{belief-propagation} and originally put forward in
%the computer science community, have been proved to be
%equivalent to the solution of the CVM equations at the
%Bethe-Peierls level. As a consequence, CVM concepts have been
%used to improve the accuracy of such algorithms, thereby
%denoted as \emph{generalized} belief propagation.

\section{Cluster-variational approach}

\label{sec:cluster-variational_approach}

\subsection{The reference model}

\label{subsec:the_reference_model}

As mentioned above, we consider quite a generic model, made up
of discrete random variables ${\xi_{i}}^{(t)}$, associated with
the vertices (or \emph{sites}) ${i=1,\dots,N}$ of a given
undirected graph, and depending on a discrete time index
${t=0,\dots,\tau}$, where $\tau$ may possibly tend to infinity.
This collection of random variables may be regarded as a
multivariate stochastic process ${\xi}^{(t)} \equiv
[{\xi_{1}}^{(t)},\dots,{\xi_{N}}^{(t)}]$, whereas a particular
realization of such a process ${x^{(0)}\!,\dots,x^{(\tau)}}$
may be denoted as a \emph{path}. Accordingly, we define the
\emph{path probability} as the probability of each given path,
namely
%\begin{equation}
%  {\wp}\,(x^{(0)}\!,x^{(1)}\!,x^{(2)}\!,\dots)
%  \triangleq
%  \mathbb{P} \{ {\xi}^{(0)} = x^{(0)} ,\, {\xi}^{(1)} = x^{(1)} ,\, {\xi}^{(2)} = x^{(2)} ,\, \dots \}
%  \, .
%\end{equation}
\begin{equation}
  {\wp}\,(x^{(0)}\!,\dots,x^{(\tau)})
  \triangleq
  \mathbb{P} \{ {\xi}^{(0)} = x^{(0)} ,\, \dots ,\, {\xi}^{(\tau)} = x^{(\tau)} \}
  \, .
\end{equation}
In the assumption of a Markovian dynamics, the path probability
takes the general form
\begin{equation}
  {\wp}\,(x^{(0)}\!,\dots,x^{(\tau)})
  = {p}^{(0)}(x^{(0)})
  \prod_{t=0}^{\tau-1} {w}^{(t)}(x^{(t+1)}|\,x^{(t)})
  \, , \label{eq:path_probability}
\end{equation}
where ${p}^{(0)}(x^{(0)})$ is the initial condition (i.e., the
probability of the initial configuration $x^{(0)}$) and
${w}^{(t)}(x^{(t+1)}|\,x^{(t)})$ are the transition
probabilities from $x^{(t)}$ to $x^{(t+1)}$ (i.e., the
conditional probability of the latter configuration at time
${t+1}$, given the former at time ${t}$).

The graph structure defines which variables in the
``current-time'' configuration are actually relevant to
determine the ``next-time'' configuration, and therefore the
structure of the transition probabilities. In many physical
models, such as those mentioned in the introduction, the
next-time configuration of each given site depends only on the
current-time configuration of its neighbors and also possibly
of the same site. In formulae, this statement means that the
transition probabilities can be written in the following
factorized form
\begin{equation}
  {w}^{(t)}(y|x)
  = \prod_{i} {{w}_{i}}^{(t)}(y_{i}|x_{i,\partial i})
  \, , \label{eq:transition_probabilities}
\end{equation}
where the product runs over all sites, $x$ and $y$
(respectively, $x_{i}$ and $y_{i}$) denote two generic
configurations of the whole graph (respectively, of the $i$
site)\footnote{If possible, throughout the paper we shall
denote a current-time configuration by a letter $x$ and a
next-time configuration by a letter $y$.}, and $x_{i,\partial
i}$ is a shorthand for a list of configuration variables
$x_{i},\{x_{j}\}_{j \in
\partial i}$ (i.e., those associated with the $i$ site and its
neighbors). We shall also assume that the initial
configurations of different sites are statistically
independent, namely that
\begin{equation}
  {p}^{(0)}(x)
  = \prod_{i} {{p}_{i}}^{(0)}(x_{i})
  \, . \label{eq:initial_conditions}
\end{equation}
The latter assumption could also be partially relaxed, but
usually there is no specific interest in doing so.

\subsection{Thermodynamic formalism and cluster variational method}

\label{subsec:thermodynamic_formalism}

Let us now consider the following variational functional
%\begin{equation}
%  \mathcal{F} [{\wp}] =
%  \sum_{x^{(0)}\!,x^{(1)}\!,x^{(2)}\!,\dots}
%  {\wp}\,(x^{(0)}\!,x^{(1)}\!,x^{(2)}\!,\dots)
%  \ln \frac{
%  {\wp}\,(x^{(0)}\!,x^{(1)}\!,x^{(2)}\!,\dots)
%  }{
%  {p}^{(0)}(x^{(0)})
%  \prod_{t=0,1,2,\dots} {w}^{(t)}(x^{(t+1)}|\,x^{(t)})
%  }
%  \, .
%\end{equation}
\begin{equation}
  \mathcal{F} [{\wp}] =
  \sum_{x^{(0)}\!,\dots,x^{(\tau)}}
  {\wp}\,(x^{(0)}\!,\dots,x^{(\tau)})
  \ln \frac{
  {\wp}\,(x^{(0)}\!,\dots,x^{(\tau)})
  }{
  {p}^{(0)}(x^{(0)})
  \prod_{t=0}^{\tau-1} {w}^{(t)}(x^{(t+1)}|\,x^{(t)})
  }
  \, ,
\end{equation}
where the sum runs over all possible paths, the path
probability $\wp$ is regarded as a variational parameter, and
the initial distribution ${p}^{(0)}$ and the transition
probabilities ${w}^{(t)}$ are assumed to be ``input
parameters'', incorporating our knowledge of the stochastic
dynamics. From the mathematical point of view, this functional
is a so-called \emph{Kullback-Leibler divergence} between the
two sides of equation \eref{eq:path_probability}, and is known
to have an absolute minimum (zero) where the two distributions
are equal, that is when equation \eref{eq:path_probability} is
verified. Introducing the Shannon information entropy
associated with the path probability, namely
%\begin{equation}
%  \mathcal{S} [{\wp}] = -
%  \sum_{x^{(0)}\!,x^{(1)}\!,x^{(2)}\!,\dots}
%  {\wp}\,(x^{(0)}\!,x^{(1)}\!,x^{(2)}\!,\dots)
%  \ln
%  {\wp}\,(x^{(0)}\!,x^{(1)}\!,x^{(2)}\!,\dots)
%  \, .
%\end{equation}
\begin{equation}
  \mathcal{S} [{\wp}] = -
  \sum_{x^{(0)}\!,\dots,x^{(\tau)}}
  {\wp}\,(x^{(0)}\!,\dots,x^{(\tau)})
  \ln
  {\wp}\,(x^{(0)}\!,\dots,x^{(\tau)})
  \, ,
\end{equation}
we can rewrite the variational functional as
\begin{equation}
  \mathcal{F} [{\wp}] =
  \sum_{x^{(0)}\!,\dots,x^{(\tau)}}
  \wp\,(x^{(0)}\!,\dots,x^{(\tau)}) \,
  \varepsilon\,(x^{(0)}\!,\dots,x^{(\tau)})
  - \mathcal{S}[{\wp}]
  \, , \label{eq:free_energy}
\end{equation}
where we have also defined
\begin{equation}
  \varepsilon\,(x^{(0)}\!,\dots,x^{(\tau)}) \triangleq
  - \ln {p}^{(0)}(x^{(0)})
  - \sum_{t=0}^{\tau-1} \ln {w}^{(t)}(x^{(t+1)}|\,x^{(t)})
  \, . \label{eq:hamiltonian}
\end{equation}
Equation \eref{eq:free_energy} points out that the variational
functional $\mathcal{F}[\wp]$ is analogous to a free energy for
a system in which time plays the role of an extra dimension,
and whose energy function $\varepsilon$ is defined by
\eref{eq:hamiltonian}. More precisely, such a system can be
viewed as a ``stack'' of ${\tau+1}$ ``copies'' of the original
graph, each one labeled by a time index ${t = 0,1,\dots,\tau}$,
so that we shall denote it as a \emph{space-time thermodynamic
system}. The path probability minimizing $\mathcal{F}[\wp]$ is
analogous to a Boltzmann distribution for the space-time
thermodynamic system.

Up to this point, we have got just a reformulation of the
problem, which is of no practical use in a fully general case.
Nevertheless, our assumptions
\eref{eq:transition_probabilities} and
\eref{eq:initial_conditions} about, respectively, transition
probabilities and initial conditions allow us to rewrite the
energy function as
\begin{equation}
  \fl
  \varepsilon\,(x^{(0)}\!,\dots,x^{(\tau)}) =
  - \sum_{i} \ln {{p}_{i}}^{(0)}({x_{i}}^{(0)})
  - \sum_{t=0}^{\tau-1} \sum_{i} \ln {{w}_{i}}^{(t)}({x_{i}}^{(t+1)}|\,{x_{i,\partial i}}^{(t)})
  \, . \label{eq:hamiltonian2}
\end{equation}
The latter expression points out that (the logarithm of) the
elementary factors of the transition probabilities and of the
initial conditions play respectively the role of interaction
energies and external fields, and that both kinds of term are
\emph{local} in both space and time. As a consequence, since
the energetic term of the free energy \eref{eq:free_energy} is
local, we expect that the (inherently nonlocal) entropic term
could be effectively approximated by a truncation of its
\emph{cumulant expansion}, according to Kikuchi's
cluster-variational method~\cite{Pelizzola2005,An1988}. Each
cumulant is associated with a given \emph{cluster} (i.e., a
given subset of lattice sites), so that different levels of
approximation may be defined by the set of \emph{maximal
clusters} retained in the expansion. The approximate entropy
can be finally recast into a linear combination of
\emph{cluster entropies}, i.e., information entropies of
\emph{cluster distributions} (the latter being obviously
defined as joint probability distributions for the random
variables associated with all sites in the given cluster).
Thus, a generic CVM entropy takes the following
form~\cite{Pelizzola2005,An1988}
\begin{equation}
  \mathcal{S}_\mathrm{CVM} [ \{{{\wp}_{\alpha}}\}_{\alpha \in \mathsf{R}} ]
  = \sum_{\alpha \in \mathsf{R}} a_{\alpha} \mathcal{S}[{{\wp}_{\alpha}}]
  \, , \label{eq:CVM-entropy}
\end{equation}
where the sum runs over the set $\mathsf{R}$ of all maximal
clusters and their subclusters, ${{\wp}_{\alpha}}$ denotes the
cluster distribution for cluster $\alpha$ (i.e., a particular
marginal of $\wp$), and $\mathcal{S}[{{\wp}_{\alpha}}]$ is the
corresponding information entropy. The coefficients
$a_{\alpha}$ of the combination, usually denoted as
\emph{M\"obius
numbers}~\cite{Pelizzola2005,HeskesAlbersKappen2003}, or
\emph{(over)counting
numbers}~\cite{YedidiaFreemanWeiss2005,HeskesAlbersKappen2003},
depend only on the graph geometry and the choice of maximal
clusters. They can be determined by a simple sum rule, which
can be stated as follows~\cite{An1988}
\begin{equation}
  \sum_{\beta \in \mathsf{R} \, | \, \beta \supseteq \alpha} a_{\beta} = 1
  \qquad \forall \alpha \in \mathsf{R}
  \, , \label{eq:sum_rule}
\end{equation}
where $\alpha$ may be any maximal cluster or subcluster, and
the sum runs over all (maximal clusters or subclusters) $\beta$
containing $\alpha$. The rule shows in particular that the
coefficient of any maximal cluster is equal to $1$, because in
that case the sum contains only one term.

Let us finally note that the choice of maximal clusters is
generally driven by physical intuition, since it defines the
range of correlations that a given approximation will be able
to take into account. Therefore, in the framework of a
space-time system like the one we are interested in, maximal
clusters should necessarily extend over different time steps,
in order to take into account time correlations. Kikuchi's
\emph{path probability method} (PPM) can be regarded, in this
framework, as a particular way of choosing the maximal
clusters, namely, one first chooses maximal clusters for a
one-time system, then every maximal cluster of the space-time
system is the union of a maximal cluster at time $t$ with its
translation at time ${t+1}$.  As already noticed in reference
\cite{Pelizzola2013}, other criteria for this choice are
possible, and may be more convenient, depending on the
structure of the transition probabilities. In the following we
shall see on test models that indeed one can obtain a
considerably better tradeoff between accuracy and computational
complexity, with respect to the conventional PPM choice.

\subsection{Specific approximations}

\label{subsec:specific_approximations}

In this section we introduce the specific approximations (i.e.,
the different choices of maximal clusters), whose performance
is analyzed in the following. First of all, let us define
precisely the types of clusters (either maximal or not) that
appear in at least one of the CVM entropies considered. Looking
at table~\ref{tab:clusters} we see that each cluster extends
over either a single timestep or two consecutive ones. Of
course, since we assume that the graph does not change with
time, clusters at different timesteps are simply translations
of one another. We introduce specific symbols (letters) to
denote the different cluster types and the related probability
distributions; such symbols are defined in
table~\ref{tab:clusters} as well.

\begin{table}
  \caption{The 1st column displays cluster type identifiers.
  The 2nd and 3rd columns report lists of sites that belong to each cluster at timesteps $t$ and ${t+1}$,
  respectively (as usual ${\partial i}$ denotes all the nearest neighbors of $i$, whereas $i,j$ denote a nearest neighbor pair).
  The 4th column displays the definition of the probability distribution function (pdf) of each cluster.
  Of course ${\xi_{i,j}}^{(t)}$ is a shorthand for ${{\xi_{i}}^{(t)}\!,{\xi_{j}}^{(t)}}$, whereas
  ${\xi_{i,\partial i}}^{(t)}$ is a shorthand for ${{\xi_{i}}^{(t)}\!,\{{\xi_{j}}^{(t)}\}_{j \in \partial i}}$
  (and the same for ${t+1}$). Analogous shorthands are used for the ``dummy'' variables $x$ and $y$.}
  \label{tab:clusters}
\flushright
\begin{tabular}{|c||l|l|lcl|}
  \hline
  % after \\: \hline or \cline{col1-col2} \cline{col3-col4} ...
  & $t$ & $t+1$ & pdf & & definition\\
  \hline \hline
  M & $i,\partial i$ & $i,\partial i$ & ${M_{i}}^{(t)}(y_{i,\partial i},x_{i,\partial i})$ & $\triangleq$ & $\mathbb{P} \{ {\xi_{i,\partial i}}^{(t+1)} = y_{i,\partial i} ,\, {\xi_{i,\partial i}}^{(t)} = x_{i,\partial i} \}$ \\
  P & $i,\partial i$ & $i           $ & ${P_{i}}^{(t)}(y_{i},x_{i,\partial i})$            & $\triangleq$ & $\mathbb{P} \{ {\xi_{i}}^{(t+1)} = y_{i} ,\, {\xi_{i,\partial i}}^{(t)} = x_{i,\partial i} \}$ \\
  Q & $i,j         $ & $i,j         $ & ${Q_{ij}}^{(t)}(y_{i,j},x_{i,j})$                  & $\triangleq$ & $\mathbb{P} \{ {\xi_{i,j}}^{(t+1)} = y_{i,j} ,\, {\xi_{i,j}}^{(t)} = x_{i,j} \}$ \\
  R & $i           $ & $i,\partial i$ & ${R_{i}}^{(t)}(y_{i,\partial i},x_{i})$            & $\triangleq$ & $\mathbb{P} \{ {\xi_{i,\partial i}}^{(t+1)} = y_{i,\partial i} ,\, {\xi_{i}}^{(t)} = x_{i} \}$ \\
  S & $i,\partial i$ & --             & ${S_{i}}^{(t)}(x_{i,\partial i})$                  & $\triangleq$ & $\mathbb{P} \{ {\xi_{i,\partial i}}^{(t)} = x_{i,\partial i} \}$ \\
  T & $i,j         $ & $i           $ & ${T_{i,ij}}^{(t)}(y_{i},x_{i,j})$                  & $\triangleq$ & $\mathbb{P} \{ {\xi_{i}}^{(t+1)} = y_{i} ,\, {\xi_{i,j}}^{(t)} = x_{i,j} \}$ \\
  U & $i           $ & $i,j         $ & ${U_{ij,i}}^{(t)}(y_{i,j},x_{i})$                  & $\triangleq$ & $\mathbb{P} \{ {\xi_{i,j}}^{(t+1)} = y_{i,j} ,\, {\xi_{i}}^{(t)} = x_{i} \}$ \\
  V & $i           $ & $i           $ & ${V_{i}}^{(t)}(y_{i},x_{i})$                       & $\triangleq$ & $\mathbb{P} \{ {\xi_{i}}^{(t+1)} = y_{i} ,\, {\xi_{i}}^{(t)} = x_{i} \}$ \\
  Z & $i,j         $ & --             & ${Z_{ij}}^{(t)}(x_{i,j})$                          & $\triangleq$ & $\mathbb{P} \{ {\xi_{i,j}}^{(t)} = x_{i,j} \}$ \\
  A & $i           $ & --             & ${A_{i}}^{(t)}(x_{i})$                             & $\triangleq$ & $\mathbb{P} \{ {\xi_{i}}^{(t)} = x_{i} \}$ \\
  \hline
\end{tabular}
\end{table}

As usual for the CVM, the simplest choice of maximal clusters
is such that the latter coincide with the groups of variables
coupled by the elementary interaction terms appearing in the
energy function. Looking at equation \eref{eq:hamiltonian2}, it
is evident that these clusters coincides with the P-clusters,
defined in table~\ref{tab:clusters}, for every
${t=0,\dots,\tau-1}$. We shall call this the P~approximation.
This approximation can be improved by including in the set of
maximal clusters also the Q-clusters, which are expected to
take into account the time-correlation of two
(spatially-correlated) nearest neighbors. This will be denoted
as the PQ approximation. A further improvement can be achieved
by adding also the R-clusters, which are expected to better
take into account the correlations among all the neighbors of a
given site (at the same timestep), because all of them are
influenced by the configuration of the ``central site'' at the
previous timestep. The latter is the PQR approximation.
Finally, looking at table~\ref{tab:clusters}, we recognize that
every P, Q, and R cluster is contained in at least one M
cluster, so that choosing the set of all M clusters as maximal
clusters may lead to an even more accurate approximation (M
approximation). Note that the M approximation is the only
choice (among those considered here), which meets Kikuchi's PPM
criterion (as stated in the previous subsection), and that
therefore can be regarded as  an instance of PPM. Note also
that it is the \emph{minimal} PPM instance for the current
reference model, because the M clusters are the minimal
clusters, which simultaneously meet Kikuchi's criterion and
``contain'' the elementary interaction terms.

\begin{table}
  \caption{M\"obius numbers for the specific approximations considered.
  Each row corresponds to a different approximation (i.e., a different choice of maximal clusters),
  whereas each column corresponds to a different cluster type.
  The numbers are usually the same for all
  $t=0,\dots,\tau-1$; certain cases in which the numbers are zero for ${t=0}$
  are indicated by an extra $0$ in parentheses.
  ${d_i \triangleq |\partial i|}$ denotes the number of neighbors
  (a.k.a. coordination number, connectivity, or degree) of a node $i$.}
  \label{tab:superposition_coefficients}
\flushright
\begin{tabular}{|l||c|r|r|r|r|r|r|r|r|r|}
  \hline
  % after \\: \hline or \cline{col1-col2} \cline{col3-col4} ...
  % \hphantom{approximation} clusters
  & M & P & Q & R & S$\ \hphantom{(0)}$ & T & U & V & Z$\ \hphantom{(0)}$ & A$\ \hphantom{(0)}$ \\
  % approximations & & & & & & & & & & & & & \\
  \hline
  \hline
  P   & $0$ & $1$ & $ 0$ & $0$ & $ 0 \ \hphantom{(0)}$ & $ 0$ & $ 0$ & $      0$ & $-1 \ \hphantom{(0)}$ & $     -1 \ (0)$            \\
  \hline
  PQ  & $0$ & $1$ & $ 1$ & $0$ & $ 0 \ \hphantom{(0)}$ & $-1$ & $ 0$ & $      0$ & $-1 \ (0)$            & $d_{i}-1 \ (0)$            \\
  \hline
  PQR & $0$ & $1$ & $ 1$ & $1$ & $-1 \ (0)$            & $-1$ & $-1$ & $d_{i}-1$ & $ 1 \ (0)$            & $      0 \ \hphantom{(0)}$ \\
  \hline
  M   & $1$ & $0$ & $-1$ & $0$ & $-1 \ (0)$            & $ 0$ & $ 0$ & $      0$ & $ 1 \ (0)$            & $      0 \ \hphantom{(0)}$ \\
  \hline
\end{tabular}
\end{table}

Making use of the sum rule \eref{eq:sum_rule}, one can
determine the M\"obius numbers, and therefore the approximate
entropy functional, for each specific approximation. This is a
rather standard calculation, whose results are summarized in
table~\ref{tab:superposition_coefficients}. Note that, as
previously mentioned, all (though not only) the M\"obius
numbers associated with the maximal clusters are equal to
unity. Note also that, in the M approximation, P, Q, and R
clusters are not maximal, so their M\"obius numbers do not need
to be (and in fact they are not) equal to unity. In the
remainder of this section we report the entropy functionals in
formulae. As far as the P approximation is concerned, we have
%\begin{equation}
%  \mathcal{S}_\mathrm{P} [ P,Z,A ]
%  = \sum_{i,t} \left( \mathcal{S}[{P_{i}}^{(t)}] - \mathcal{S}[{A_{i}}^{(t)}] \right)
%  - \sum_{ij,t} \mathcal{S}[{Z_{ij}}^{(t)}]
%  \, .
%\end{equation}
\begin{equation}
  \mathcal{S}_\mathrm{P} [ P,Z,A ]
  = \sum_{t=0}^{\tau-1}
  \left\{
  \sum_{i} \mathcal{S}[{P_{i}}^{(t)}]
  - \sum_{ij} \mathcal{S}[{Z_{ij}}^{(t)}]
  \right\}
  - \sum_{t=1}^{\tau-1}
  \sum_{i} \mathcal{S}[{A_{i}}^{(t)}]
  \, ,
\end{equation}
where $\sum_{i}$ and $\sum_{ij}$ denote sums over all sites and
over all nearest neighbor pairs, respectively, and
${\mathcal{S}[\cdot]}$ denotes the Shannon entropy of each
given probability distribution. The arguments ${P,Z,A}$ mean
that the functional $\mathcal{S}_\mathrm{P}$ depend on all P-,
Z-, and A-type cluster distributions, i.e., $P$ is a shorthand
for $\{{P_{i}}^{(t)}: \forall t=0,1,\dots,\tau-1, \forall i\}$,
and so on. With analogous notations, the PQ entropy
approximation reads
\begin{eqnarray}
  \fl
  \mathcal{S}_\mathrm{PQ} [ P,Q,T,Z,A ]
  & = \sum_{t=0}^{\tau-1}
  \left\{
  \sum_{i} \mathcal{S}[{P_{i}}^{(t)}]
  + \sum_{ij} \mathcal{S}[{Q_{ij}}^{(t)}]
  - \sum_{i,j \in \partial i} \mathcal{S}[{T_{i,ij}}^{(t)}]
  \right\}
  + \nonumber \\ &
  - \sum_{t=1}^{\tau-1}
  \left\{
  \sum_{ij} \mathcal{S}[{Z_{ij}}^{(t)}]
  - \sum_{i} (d_{i}-1) \mathcal{S}[{A_{i}}^{(t)}]
  \right\}
  \, ,
\end{eqnarray}
where $\sum_{i,j \in \partial i}$ denotes a double sum over all
sites and over all neighbors of each given site. Furthermore,
the PQR entropy reads
\begin{eqnarray}
  \fl
  \mathcal{S}_\mathrm{PQR} [ P,Q,R,S,T,U,V,Z ]
  = \sum_{t=0}^{\tau-1}
  \left\{
  \sum_{i}
  \left[
  \mathcal{S}[{P_{i}}^{(t)}]
  + \mathcal{S}[{R_{i}}^{(t)}]
  + (d_{i}-1) \mathcal{S}[{V_{i}}^{(t)}]
  \right]
  + \right. \nonumber \\ \fl \left.
  + \sum_{ij} \mathcal{S}[{Q_{ij}}^{(t)}]
  - \sum_{i,j \in \partial i}
  \left[
  \mathcal{S}[{T_{i,ij}}^{(t)}]
  + \mathcal{S}[{U_{ij,i}}^{(t)}]
  \right]
  \right\}
%  + \nonumber \\ &
  - \sum_{t=1}^{\tau-1}
  \left\{
  \sum_{i} \mathcal{S}[{S_{i}}^{(t)}]
  - \sum_{ij} \mathcal{S}[{Z_{ij}}^{(t)}]
  \right\}
  , \nonumber \\ \label{eq:entropia-PQR}
\end{eqnarray}
whereas the M entropy reads
\begin{eqnarray}
  \fl
  \mathcal{S}_\mathrm{M} [ M,Q,S,Z ]
  = \sum_{t=0}^{\tau-1}
  \! \left\{
  \sum_{i} \mathcal{S}[{M_{i}}^{(t)}]
  - \sum_{ij} \mathcal{S}[{Q_{ij}}^{(t)}]
  \right\} \!
  - \sum_{t=1}^{\tau-1}
  \! \left\{
  \sum_{i} \mathcal{S}[{S_{i}}^{(t)}]
  - \sum_{ij} \mathcal{S}[{Z_{ij}}^{(t)}]
  \right\} \!
  . \nonumber \\
\end{eqnarray}
For each approximation, the free-energy functional is obtained
from equation \eref{eq:free_energy} by replacing the entropic
term $\mathcal{S}$ with one of the approximate entropies. As
previously mentioned, the energetic term does not require any
further approximation, because the maximal clusters are
properly chosen to ``contain'' the elementary coupling terms of
the energy function \eref{eq:hamiltonian}.

The minimization of the resulting functionals is far from being
a straightforward problem. This is ultimately due to cluster
overlapping, which entails that compatibility constraints
between cluster distributions must be satisfied. In general,
the constrained minimization problem for a generic CVM free
energy can be solved iteratively by a message-passing procedure
denoted as \emph{generalized belief
propagation}~\cite{Pelizzola2005,YedidiaFreemanWeiss2005}
(which is not guaranteed to converge), or by more complex
(provably convergent) methods~\cite{HeskesAlbersKappen2003}.
Nevertheless, the peculiar structure of the couplings, arising
in a space-time system describing a Markovian stochastic
dynamics like that of our reference model, allows one to devise
a much simpler procedure, which does not require iterative
refinement, but only a ``natural'' sequence of iterations,
starting from the initial conditions and following the
(probabilistic) time evolution of the system. We report the
details of such a procedure in \ref{app:details}, only for the
PQR approximation. The differences occurring in the other
approximations (which are indeed very much analogous) are
discussed in \ref{app:other_approximations}. Furthermore,
analogous procedures can be used to investigate continuous-time
processes (in a discretized form), and, as mentioned in the
introduction, in such a case one can argue that the PQR and PQ
approximations become equivalent. The latter issues are
discussed in \ref{app:continuous_time}.

\subsection{Related works}

\label{subsec:related_works}

Before switching to a performance evaluation of the different
approximations described above, we find it worth summarizing a
number of connections with previous works, which emerge from
our analysis. As mentioned in the introduction, in most cases
equivalent approximations have been put forward without
resorting to a variational scheme, but just on the basis of
heuristic reasoning, and/or for specific model dynamics. In
such cases, our approach retains some interest in that it
offers a more general and unified view of the matter.

First of all, it is possible to recognize that the P and PQ
schemes turn out to be respectively equivalent to the classical
(non-variational) \emph{mean-field} and \emph{pairwise}
approximations, which have been used, for instance, to
investigate epidemic spreading phenomena on
networks~\cite{WangTangStanleyBraunstein2016,PastorsatorrasCastellanoVanmieghemVespignani2015,MataFerreira2013}.
In such a context, the P approximation is customarily denoted
as \emph{quenched mean field}, in order to distinguish it from
a naive mean-field approach, in which the infection probability
is assumed to be a node-independent
quantity~\cite{WangTangStanleyBraunstein2016}, and from the
popular \emph{heterogeneous mean-field approach} by
Pastor-Satorras and
Vespignani~\cite{PastorsatorrasVespignani2001}, in which nodes
are distinguished only on the basis of their degree. For
epidemics on networks that are expected to be inherently
homogeneous, such as random-regular graphs or regular lattices,
the same approximations have been considered also by Petermann
and De Los Rios~\cite{PetermannDelosrios2004}, and other
authors~\cite{MatsudaOgitaSasakiSato1992}. We point out these
equivalences in \ref{subapp:PQandP} along with some details
about the respective iterative procedures. Secondly, the M
approximation turns out to be an instance of Kikuchi's
(variational) path-probability method, as we have mentioned in
the previous subsection, but it has also been derived
independently (for the case of homogeneous epidemics) again by
Petermann and De Los Rios~\cite{PetermannDelosrios2004}, who
denoted it as \emph{star approximation}. The latter equivalence
is detailed in \ref{subapp:M}. Furthermore, let us note that,
in the special case of a spatially one-dimensional system
(i.e., when the graph is a linear chain), a recent analysis of
nonlinear Voter models carried out by Schweitzer and
Behera~\cite{SchweitzerBehera2015} makes use of two different
approximation schemes (denoted as \emph{triplet} and
\emph{quintuplet} approximations), which turn out to be special
cases of the PQ and M approximations, respectively.

Another special case of our reference model, which nonetheless
includes several models of physical relevance (such as the
kinetic Ising model, or the ordinary Voter model), is that in
which the elementary transition probability for a given site
$i$ (namely, ${w_{i}}^{(t)}(y_{i}|x_{i, \partial i})$) does not
depend on the current-time configuration of the site itself
$x_{i}$ (so that we can denote it as
${w_{i}}^{(t)}(y_{i}|x_{\partial i})$). In other words, the
next-time configuration of each site is conditioned only by the
current-time configurations of its neighbors. For this reason,
we shall denote this case as a purely
\emph{neighbor-conditioned} dynamics. As mentioned in the
introduction, the \emph{star} and \emph{diamond} approximations
of reference~\cite{Pelizzola2013} refer to this kind of
dynamics. Actually, it turns out that, for a generic
purely-neighbor-conditioned dynamics, the PQ and P
approximations degenerate into each other, and both coincide
with the star approximation of reference~\cite{Pelizzola2013}.
Similarly, the M and PQR approximations degenerate into each
other, and coincide with the diamond approximation. All these
connections are detailed in \ref{subapp:diamond} and
\ref{app:other_approximations}. As a further consequence, we
can state that in this case the diamond approximation is itself
an instance of Kikuchi's path probability method.

\section{Test models}

\label{sec:test_models}

In this section we test our approximations against Monte Carlo
simulations on specific instances of a SIS epidemic model on
(sparse) graphs drawn from well known random ensembles, namely
the Erd\"os-R\'enyi ensemble and the random-regular ensemble.
In the former case, the degree distribution is poissonian,
whereas in the latter case all vertices have the same degree,
so that we also denote the two kinds of graphs as
\emph{poissonian} and \emph{uniform}, respectively.

We consider a SIS model with parallel dynamics, whose
definition is as follows. Each site variable ${\xi_{i}}^{(t)}$
can take two possible values, ${x_{i}=0,1}$, meaning that the
$i$ site is susceptible or infected, respectively. An infected
site $i$ has a probability $\mu_{i}$ to recover (i.e., to
become susceptible) at the next timestep. Moreover, a
susceptible site $i$ has a probability $\beta_{j \to i}$ to
become infected because of any of its neighbors $j$, if the
latter is infected at the current timestep, and $0$ otherwise.
In other words, the probability of an infection event from $j$
to $i$ is ${x_{j} \beta_{j \to i}}$. Since infection events by
different neighbors are not mutually exclusive, the \emph{total
probability} of a given site $i$ to become infected can be more
easily computed via its complementary probability (i.e., the
probability that $i$ remains susceptible), which is equal to
the probability that none of the neighbors transmits the
contagion. Assuming independence among different infection
events, the transition probability to a susceptible state can
be finally written as
\begin{equation}
  w_{i}(0|x_{i, \partial i})
  = x_{i} \mu_{i}
  + (1 - x_{i}) \prod_{j \in \partial i} \left( 1 - x_{j} \beta_{j \to i} \right)
  \, ,
\end{equation}
whereas, by normalization, the transition probability to an
infected state will be
\begin{equation}
  w_{i}(1|x_{i, \partial i})
  = 1 - w_{i}(0|x_{i, \partial i})
  \, .
\end{equation}

We consider two fixed graphs with ${N=1000}$ vertices, a
poissonian one with average degree ${c = 4}$ and a uniform one
with degree ${c = 4}$. As far as model parameters are
concerned, in both cases we choose a uniform recovery
probability, namely ${\mu_{i} = \mu = 0.5}$, whereas the
infection probability is chosen to be homogeneous in the former
case (namely, ${\beta_{j \to i} = \beta = \mu/3}$) and
heterogeneous in the latter (namely, $\beta_{j \to i}$ drawn at
random with uniform distribution between $0$ and $\mu$). Note
that, as a result, both systems turn out to be heterogeneous:
in the Erd\"os-R\'enyi system, heterogeneity arises from the
random graph structure, whereas, in the random-regular system,
heterogeneity arises from the randomness of infection
probabilities. The initial conditions are chosen in such a way
that both systems have on average 1\% of infected sites, but
with a slight difference. In the poissonian case, we use
``quenched'' initial conditions, that is, 1\% of (randomly
drawn) sites are assigned the initial distribution
\begin{equation}
  {p_{i}}^{(0)}(x_{i}) = x_{i}
  \, ,
\end{equation}
whereas the remaining 99\% are assigned the distribution
\begin{equation}
  {p_{i}}^{(0)}(x_{i}) = 1-x_{i}
  \, .
\end{equation}
In the uniform case we use ``annealed'' initial conditions,
that is, all sites are assigned the distribution
\begin{equation}
  {p_{i}}^{(0)}(x_{i}) = 0.01 \, x_{i} + 0.99 \, (1 - x_{i})
  \, .
\end{equation}
In the simulation, quenched conditions correspond to all
realizations initialized with exactly the same configuration
(with 1\% of infected sites) drawn at random at the beginning,
whereas annealed conditions correspond to a different initial
configuration (still with 1\% of infected sites) drawn at
random for each realization. Note that there is no specific
reason for associating a particular type of initial condition
with a particular type of graph. We simply observe that
different initial conditions give rise to a slightly different
(probabilistic) short-time behavior of the system, appearing
independently of the graph type, and we are meant to analyze
whether such a behavior is more or less accurately reproduced
by the approximations.

\begin{figure}
  \flushright \resizebox{132mm}{!}{\includegraphics*{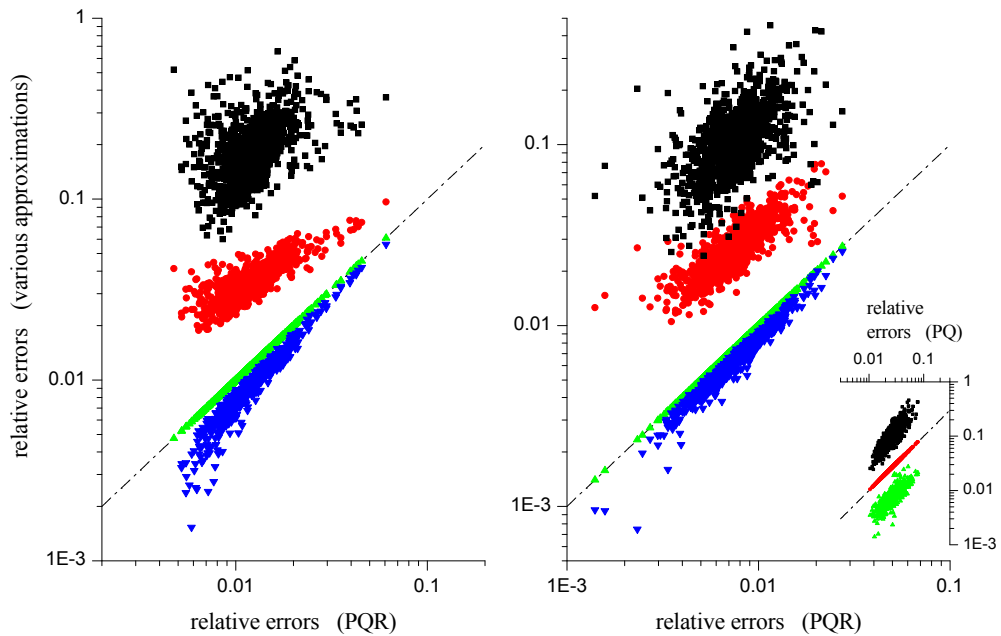}}
  \caption
  {
    Relative errors between theory and simulations
    for the steady-state infection probability of individual sites (see the text).
    The left and right panels refer to the poissonian and uniform graph cases, respectively.
    Data obtained by different approximations are plotted (with different colors and symbols)
    as a function of data obtained by the PQR approximation
    (black squares = P approximation, red circles = PQ approximation,
    green up-triangles = PQR approximation, blue down-triangles = M approximation).
    In the right-panel inset, various data (same color/symbol convention)
    are plotted as a function of the PQ data.
    Statistical uncertainties are smaller than the symbol thickness.
  }
  \label{fig:ss}
\end{figure}

For each approximation, each graph and parameters choice, and
each site ${i=1,\dots,N}$, we can compute the probability of
the infected state (infection probability) as a function of
time, namely
\begin{equation}
  {\rho_{i}}^{(t)}
  \triangleq
  \mathbb{E} \{ {\xi_{i}}^{(t)} \}
  = \mathbb{P} \{ {\xi_{i}}^{(t)} = 1 \}
  = {A_{i}}^{(t)}(1)
  \, ,
\end{equation}
where the single-site (A-cluster) distribution
${A_{i}}^{(t)}(x_{i})$ can be determined as a marginal of a
maximal-cluster distribution (for instance the P cluster):
\begin{equation}
  {A_{i}}^{(t)}(x_{i})
  = \sum_{x_{\partial i}} \sum_{y_{i}} {P_{i}}^{(t)}(y_{i},x_{i,\partial i})
  \, .
\end{equation}
The maximal cluster distributions are determined in turn by the
minimization procedure of the free-energy functional, as
mentioned in the previous section (and detailed in
\ref{app:details}). In the simulations, the infection
probability is evaluated as an empirical average of
${\xi_{i}}^{(t)}$ over ${10000}$ realizations (let us denote it
as ${\bar{\rho}_{i}}\vphantom{\rho_{i}}^{(t)}$). The
steady-state is easily identified in the approximations, since
we observe that, as $t$ increases, ${\rho_{i}}^{(t)}$ tends to
a well-defined plateau value $\rho_{i}$. Conversely, in the
simulations ${\bar{\rho}_{i}}\vphantom{\rho_{i}}^{(t)}$
exhibits a fluctuating behavior for large $t$, so we compute
the steady-state value $\bar{\rho}_{i}$ as an average over a
large number of timesteps, which allows us to get a suitably
small statistical uncertainty. We then compute the relative
errors between theory and simulation (in the steady state) as
${|\rho_{i}/\bar{\rho}_{i} - 1|}$. We display all these data in
figure~\ref{fig:ss}, for each approximation, as a function of
the corresponding data for the PQR approximation. Note that, as
a consequence, PQR data fall on a straight line. More
importantly, this way of showing data allows us to argue at a
glance that the M approximation is \emph{systematically} (i.e.,
for all sites) better than the PQR approximation, which is in
turn systematically better than PQ. Let us also observe that,
in the poissonian graph case (left panel), the PQ and P data
are so separated that also PQ clearly turns out to be
systematically better than P. For the uniform graph case (right
panel), the boundary between the two data ``clusters'' appears
a little more fuzzy, so that one might suspect that, at least
for some sites, the P approximation could give a better
approximation than PQ. Nevertheless, plotting as a function of
the PQ data (right panel inset) shows that this is not the
case, i.e., that PQ is still \emph{systematically} better than
P. From figure~\ref{fig:ss}, one can also assess that the
accuracy improvements that one can obtain by ``upgrading'' the
approximation level from P, to PQ, to PQR are, in relative
terms, of the same order of magnitude, whereas a further
upgrade to the M approximation achieves a minor improvement.

\begin{figure}
  \flushright \resizebox{132mm}{!}{\includegraphics*{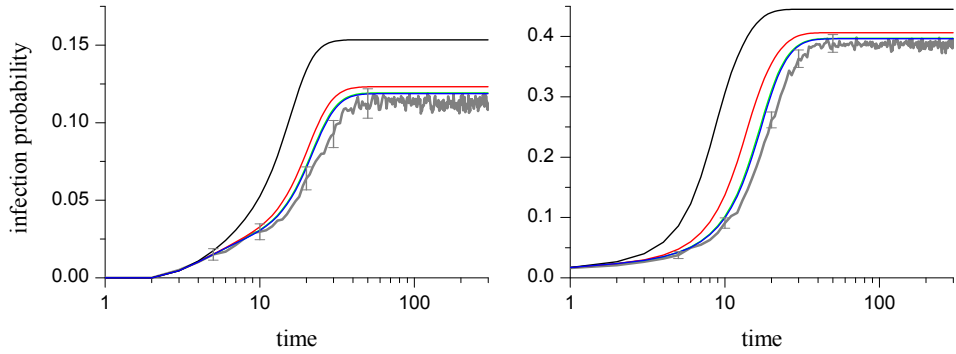}}
  \caption
  {
    Time evolution of the infection probability ${\rho_{i}}^{(t)}$
    for the worst-case site $i$ (see the text).
    The left and right panels refer to the poissonian and uniform graph cases, respectively.
    Data obtained by different approximations are plotted in different colors
    (black = P approximation, red = PQ approximation,
    green = PQR approximation, blue = M approximation).
    Simulation data are displayed as a thick grey line with error bars
    ($\pm 3$ standard deviations).
  }
  \label{fig:wc}
\end{figure}

In order to assess also the transient behavior, we display in
figure~\ref{fig:wc} the whole time evolution of the infection
probability ${\rho_{i}}^{(t)}$ for a specific site $i$. In
particular, we choose the site $i$ for which the PQR
approximation gives the largest relative error (worst case) in
the steady state. We can observe that the systematic accuracy
improvement, already noticed for the steady state, also takes
place in the short-time evolution. Such a behavior is indeed
common for all sites. We also note that the PQR and M results
are almost indistinguishable at this scale, whereas they
provide a noticeable improvement over the PQ (pair)
approximation. Moreover, we can observe that, in the case of
quenched initial conditions (left panel), the infection
probability remains sharply zero up to a given timestep (the
$i$ site being one of those initially susceptible). In such a
case, even the simplest P approximation (mean field) correctly
reproduces the very short time behavior (including the precise
timestep at which the $i$ site first exhibits a nonzero
infection probability), but otherwise it rapidly deviates from
the trend predicted by the simulations.

\begin{figure}
  \flushright \resizebox{132mm}{!}{\includegraphics*{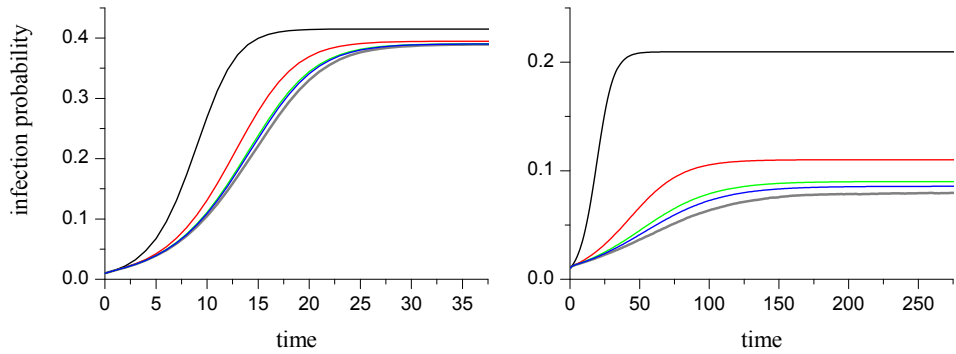}}
  \caption
  {
    Time evolution of the (node-independent) infection probability ${\rho}^{(t)}$
    for a uniform graph with uniform infection parameters (see the text).
    The left and right panels refer to the cases ${\beta = \mu/2}$ and ${\beta = \mu/3}$, respectively.
    Data obtained by different approximations and simulations are plotted in different colors,
    with the same conventions as in figure~\ref{fig:wc}.
    The statistical uncertainty of the simulation data is smaller than the line thickness.
  }
  \label{fig:uu}
\end{figure}

One more test case we consider is that of a uniform graph
(still with ${c=4}$), with homogeneous recovery probability
(${\mu_{i}=\mu=0.5}$) and, as well, homogeneous infection
probability (${\beta_{j \to i}=\beta}$). Initial conditions are
chosen to be of the annealed type (as defined above). In such a
system we expect a fully homogeneous infection probability
(i.e., ${{\rho_{i}}^{(t)} = \rho^{(t)}}$ for all $i$), so that
in the simulation we can evaluate the latter by an extra
average over the sites, thus obtaining much less noisy results.
In order to enhance the ``noise-reduction'' effect, we also
choose a larger graph size, namely ${N = 100000}$, and we
verify that, with an average over just $100$ realizations, the
statistical uncertainty turns out to be negligible at the scale
of the graphs. We display the results in figure~\ref{fig:uu}
for the specific cases ${\beta = \mu/2}$ and ${\beta = \mu/3}$.
In the former case (left panel), we can observe that almost all
the approximations (except possibly P) give a rather accurate
prediction of the steady-state infection probability. This may
be partially ascribed to the fact that in this setting the
steady state is expected to be an equilibrium state, at odds
with the previously considered cases. It is remarkable that,
even though the M, PQR, and PQ approximations are all very
accurate in the steady state, the PQ approximation shows quite
a relevant discrepancy during the transient. Furthermore, upon
decreasing $\beta$ (right panel) one can observe that the
various approximations exhibit increasingly different results
also in the steady state. This is related to the fact that the
model approaches its critical point, at which the active
(endemic) steady state disappears, and the long-time infection
probability becomes rigorously zero. In analogy with
equilibrium phase transitions, different approximations give
different predictions for the critical $\beta$ value, as
already pointed out in the continuous-time
case~\cite{PastorsatorrasCastellanoVanmieghemVespignani2015,MataFerreira2013}.

\begin{figure}
  \flushright \resizebox{132mm}{!}{\includegraphics*{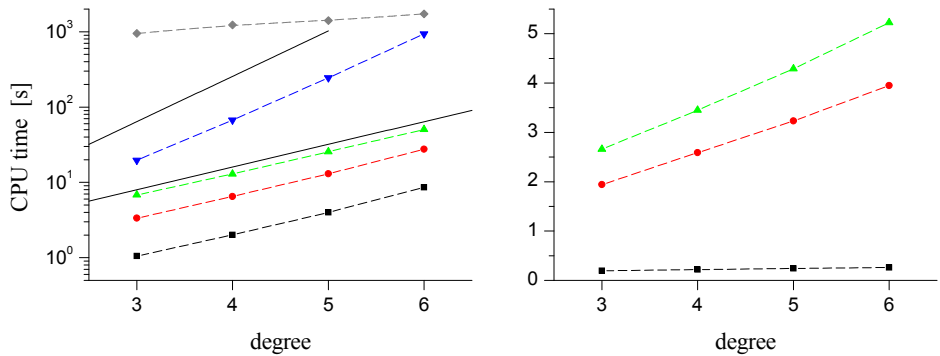}}
  \caption
  {
    CPU times (Intel Core~2 Duo) needed to
    perform $1000$ evolution timesteps of a uniform graph with ${N = 1000}$ vertices,
    as a function of the degree $c$.
    The left and right panels refer to generic and optimized programs, respectively (see the text).
    Data referring to different approximations are denoted by different symbols and colors
    (same convention as in figure~\ref{fig:ss}),
    grey diamonds denote simulation times for $10000$ realizations.
    Dashed lines are an eyeguide. Thin black lines represent the functions $2^{c}$ and $2^{2c}$.
  }
  \label{fig:tempi}
\end{figure}

Let us finally compare the different approximations in terms of
computational complexity. In figure~\ref{fig:tempi} (left
panel) we display the CPU time (Intel Core~2 Duo) needed to
perform $1000$ evolution timesteps of a uniform graph with ${N
= 1000}$ vertices, as a function of the (fixed) degree $c$. Of
course, greater accuracy corresponds to larger computational
effort, though not with a simple proportionality. In
particular, we note that the M approximation entails a very
large increase of the complexity, in spite of the fact that its
accuracy is just slightly better than that of PQR (at least for
the specific epidemic model considered here). The scaling
behavior with respect to $N$ is obviously linear, so that we do
not need to report detailed results, whereas it turns out to be
exponential with respect to $c$. In principle, the complexity
scales with the number of configurations of the largest cluster
retained in the CVM expansion, so that the P, PQ and PQR
approximations are all exponential in $c$, whereas the M
approximation is exponential in ${2c}$. From the theoretical
point of view of computational complexity, this is an
irrelevant difference, but from the practical point of view
this means that the computational cost of the M approximation
may rapidly become similar to that of a simulation, upon
increasing the degree, as shown in figure~\ref{fig:tempi}. On
the other hand, the P, PQ and PQR approximations present an
extra computational advantage, which can be stated as follows.
If the transition probabilities can be written as a product of
terms depending each on a single neighbor (or, as well, as a
sum of a few terms of this kind), simple dynamic programming
precautions allow one to reduce the complexity scaling to be
linear in the degree. Details about this issue are beyond the
scope of the current paper, but we only mention the fact that
the transition probabilities of the SIS model satisfy the above
prescription. Execution times of the optimized programs are
reported in the right panel of figure~\ref{fig:tempi}.

\section{Conclusions}

\label{sec:conclusions}

In this paper we have performed a detailed investigation of
different types of mean-field-like approximations for
stochastic dynamics on graphs, within the unifying framework of
a cluster-variational approach. In analogy with its equilibrium
counterpart, this approach naturally opens the way to improving
the approximation, by a suitable choice of the maximal
clusters, which basically define the range of correlations that
a given approximation scheme is able to take into account. On
the one hand, we have pointed out that the simplest choices of
maximal clusters (namely, the P and PQ approximations,
according to our nomenclature) coincide respectively with the
ordinary mean-field and the pairwise approximations, previously
derived in different contexts without resorting to a
variational scheme. On the other hand, we have proposed a
slightly more complex cluster choice (PQR in our nomenclature),
which provides a generalization of the so-called diamond
approximation, to dynamics that are \emph{not} purely
neighbor-conditioned (i.e., such that the transition
probability of a given site variable may also depend on the
current configuration of the site itself). Furthermore, we have
considered an even larger cluster choice (M in our
nomenclature), which turns out to be an instance of Kikuchi's
path-probability method. We have shown explicitly that, in the
case of a purely neighbor-conditioned dynamics, the P
(mean-field) and PQ (pairwise) approximations become
equivalent, and similarly the M approximation (path-probability
method) degenerates into the PQR (generalized diamond)
approximation.

We have tested the different approximations on a
(discrete-time) epidemic model of the SIS type (i.e., with
recurrent dynamics) on random graphs, possibly with
heterogeneous infection parameters. We have compared the
results with Monte Carlo simulations, and demonstrated that
increasing the maximal cluster size leads to a systematic
improvement of the level of accuracy. Such an improvement can
obviously be obtained at the cost of a higher computational
complexity, but not with a simple proportionality. In
particular, the PQR approximation yields nearly the same
accuracy as the M approximation, with a considerably lower
complexity. We have also confirmed that the PQR approximation
yields very good accuracy even in the transient regime, as
previously reported for the diamond approximation in the case
of kinetic Ising
models~\cite{DominguezDelferraroRiccitersenghi2016}. Let us
observe that our choice of a discrete-time dynamics is a bit
artificial, as continuous time is the most typical setting of
epidemic models (even though with some
exceptions~\cite{Moreno_et_al2010}). In our comparative
analysis, this choice is meant to point out the difference
between the PQR and PQ approximations, which we indeed prove to
become equivalent in the continuous-time limit.

Let us finally mention an important alternative class of
approximate methods for the same kind of problems, which go
under the name of \emph{dynamic message-passing} (DMP) or the
like. Basically, these methods rely on the so-called
\emph{dynamic cavity}
equations~\cite{NeriBolle2009,DelferraroAurell2015,Delferraro2016},
which are exact on tree graphs (and therefore likely accurate
on treelike graphs), but require taking into account the whole
time evolution of each given site variable. Different
approximations over the time dimension can then be worked out
to make the problem computationally tractable. Such alternative
approximations include for instance the early ``dynamic
cavity'' approximation by Aurell and
Mahmoudi~\cite{AurellMahmoudi2011,AurellMahmoudi2011ctp,AurellMahmoudi2012},
the \emph{Markovian closure} scheme by Del Ferraro and
Aurell~\cite{DelferraroAurell2015,Delferraro2016}, and the very
refined (and computationally demanding) matrix-product
algorithm by Barthel and
coworkers~\cite{BarthelDebaccoFranz2015}. An analogous (but
apparently different) DMP approach, directly formulated for the
continuous-time SIS model, has also been put forward
recently~\cite{ShresthaScarpinoMoore2015}. Interestingly, the
latter method has been reported to perform slightly better than
the pairwise (PQ) approximation, but mainly in the stationary
state, whereas it seems to be generally less accurate in
transient states. On the other hand, the aforementioned
Markovian-closure scheme~\cite{DelferraroAurell2015} (which can
in principle be applied to any kind of dynamics falling within
our reference model) has been thoroughly
tested~\cite{DominguezDelferraroRiccitersenghi2016} on various
(Glauber) kinetic Ising models, and reported to perform
(slightly but quite systematically) worse than the diamond
(PQR) approximation. Note that, as in this case the dynamics is
purely neighbor-conditioned, the pair approximation is no
longer a good term of comparison, because it degenerates into
the ordinary mean field, as we show in \ref{subapp:PQandP}. We
have not included in our paper a detailed comparative analysis
with respect to dynamic-message-passing approaches, because, as
previously mentioned, the main focus of our work is on the
unifying framework of the cluster variational method.
Nevertheless, such an investigation might be in fact worth
being performed, so that we shall consider it as a subject for
future work.

%A similarly interesting issue to be investigated concerns
%possible connections between (seemingly) different DMP
%approaches. In particular \dots

%To the best of our knowledge, from the existing literature it
%cannot be argued whether the latter message-passing approach
%can be derived as an instance of those mentioned above or not.

\appendix

\section{Details of the PQR approximation}

\label{app:details}

In this appendix we report in detail the method we have used to
determine the minimum of the cluster-variational functional for
the case of the PQR approximation. As we shall see, this method
naturally leads to an ``iterative'' procedure (the one we have
implemented numerically), which, starting from the initial
conditions, builds up the cluster distributions at subsequent
timesteps, up to the eventual steady state. Let us first
discuss the issue of compatibility constraints, which will be
of use in the following.

\subsection{Compatibility constraints}

\label{subapp:comp}

The probability distributions for maximal clusters and
subclusters (which for the PQR approximations have been denoted
as $P,Q,R,S,T,U,V,Z$) are of course not independent variational
parameters, because they have to satisfy (besides
normalization) a number of compatibility constraints. By
\emph{compatibility} we mean that the marginalization of
probability distributions of overlapping maximal clusters must
give, for the same subcluster, the same distribution. Writing
down these conditions for all subclusters of all maximal
clusters, we obtain a sufficient (indeed redundant) set of
constraint equations. Let us group together these equations
depending on the type of maximal cluster involved. We
understand that all the equations must hold for ${t =
0,1,\dots,\tau-1}$.
\begin{trivlist}

\item P-type maximal clusters:
\begin{subequations}
\label{eq:vinc-P}
\begin{eqnarray}
  \sum_{y_{i}} {P_{i}}^{(t)}(y_{i},x_{i,\partial i})
  & = {S_{i}}^{(t)}(x_{i,\partial i})
  & \qquad \forall i
  \, , \label{eq:comp-PS} \\
  \sum_{x_{\partial i \setminus j}} \sum_{y_{i}} {P_{i}}^{(t)}(y_{i},x_{i,\partial i})
  & = {Z_{ij}}^{(t)}(x_{i,j})
  & \qquad \forall i, \forall j \in \partial i
  \, , \label{eq:comp-PZ} \\
  \sum_{x_{\partial i \setminus j}} {P_{i}}^{(t)}(y_{i},x_{i,\partial i})
  & = {T_{i,ij}}^{(t)}(y_{i},x_{i,j})
  & \qquad \forall i, \forall j \in \partial i
  \, , \label{eq:comp-PT} \\
  \sum_{x_{\partial i}} {P_{i}}^{(t)}(y_{i},x_{i,\partial i})
  & = {V_{i}}^{(t)}(y_{i},x_{i})
  & \qquad \forall i
  \, . \label{eq:comp-PV}
\end{eqnarray}
\end{subequations}

\item Q-type maximal clusters:
\begin{subequations} \label{eq:vinc-Q}
\begin{eqnarray}
  \sum_{y_{j}} {Q_{ij}}^{(t)}(y_{i,j},x_{i,j})
  & = {T_{i,ij}}^{(t)}(y_{i},x_{i,j})
  & \qquad \forall i, \forall j \in \partial i
  \, , \label{eq:comp-QT} \\
  \sum_{y_{i,j}} {Q_{ij}}^{(t)}(y_{i,j},x_{i,j})
  & = {Z_{ij}}^{(t)}(x_{i,j})
  & \qquad \forall ij
  \, , \label{eq:comp-QZ} \\
  \sum_{x_{j}} \sum_{y_{j}} {Q_{ij}}^{(t)}(y_{i,j},x_{i,j})
  & = {V_{i}}^{(t)}(y_{i},x_{i})
  & \qquad \forall i, \forall j \in \partial i
  \, , \label{eq:comp-QV} \\
  \sum_{x_{j}} {Q_{ij}}^{(t)}(y_{i,j},x_{i,j})
  & = {U_{ij,i}}^{(t)}(y_{i,j},x_{i})
  & \qquad \forall i, \forall j \in \partial i
  \, , \label{eq:comp-QU} \\
  \sum_{x_{i,j}} {Q_{ij}}^{(t)}(y_{i,j},x_{i,j})
  & = {Z_{ij}}^{(t+1)}(y_{i,j})
  & \qquad \forall ij
  \, . \label{eq:comp-QZfwd}
\end{eqnarray}
\end{subequations}

\item R-type maximal clusters:
\begin{subequations} \label{eq:vinc-R}
\begin{eqnarray}
  \sum_{y_{\partial i \setminus j}} {R_{i}}^{(t)}(y_{i,\partial i},x_{i})
  & = {U_{ij,i}}^{(t)}(y_{i,j},x_{i})
  & \qquad \forall i, \forall j \in \partial i
  \, , \label{eq:comp-RU} \\
  \sum_{y_{\partial i}} {R_{i}}^{(t)}(y_{i,\partial i},x_{i})
  & = {V_{i}}^{(t)}(y_{i},x_{i})
  & \qquad \forall i
  \, , \label{eq:comp-RV} \\
  \sum_{x_{i}} \sum_{y_{\partial i \setminus j}} {R_{i}}^{(t)}(y_{i,\partial i},x_{i})
  & = {Z_{ij}}^{(t+1)}(y_{i,j})
  & \qquad \forall i, \forall j \in \partial i
  \, , \label{eq:comp-RZfwd} \\
  \sum_{x_{i}} {R_{i}}^{(t)}(y_{i,\partial i},x_{i})
  & = {S_{i}}^{(t+1)}(y_{i,\partial i})
  & \qquad \forall i
  \, . \label{eq:comp-RSfwd}
\end{eqnarray}
\end{subequations}

\end{trivlist}
This set of conditions also entails direct compatibility
equations between subcluster distributions (for overlapping
subclusters). Let us write explicitly some of these equations
below.
\begin{trivlist}

\item From equations \eref{eq:comp-PS} and \eref{eq:comp-PZ},
    or equivalently \eref{eq:comp-RZfwd} and
    \eref{eq:comp-RSfwd}, we have
\begin{equation}
  \sum_{x_{\partial i \setminus j}} {S_{i}}^{(t)}(x_{i,\partial i})
  = {Z_{ij}}^{(t)}(x_{i,j})
  \qquad \forall i, \forall j \in \partial i
  \, . \label{eq:comp-SZ}
\end{equation}

\item From equations \eref{eq:comp-PZ} and \eref{eq:comp-PT},
    or equivalently \eref{eq:comp-QT} and \eref{eq:comp-QZ},
    we have
\begin{equation}
  \sum_{y_{i}} {T_{i,ij}}^{(t)}(y_{i},x_{i,j})
  = {Z_{ij}}^{(t)}(x_{i,j})
  \qquad \forall i, \forall j \in \partial i
  \, . \label{eq:comp-TZ}
\end{equation}

\item From equations \eref{eq:comp-PT} and \eref{eq:comp-PV},
    or equivalently \eref{eq:comp-QT} and \eref{eq:comp-QV},
    we have
\begin{equation}
  \sum_{x_{j}} {T_{i,ij}}^{(t)}(y_{i},x_{i,j})
  = {V_{i}}^{(t)}(y_{i},x_{i})
  \qquad \forall i, \forall j \in \partial i
  \, . \label{eq:comp-TV}
\end{equation}

\item From equations \eref{eq:comp-QV} and \eref{eq:comp-QU},
    or equivalently \eref{eq:comp-RU} and \eref{eq:comp-RV},
    we have
\begin{equation}
  \sum_{y_{j}} {U_{ij,i}}^{(t)}(y_{i,j},x_{i})
  = {V_{i}}^{(t)}(y_{i},x_{i})
  \qquad \forall i, \forall j \in \partial i
  \, . \label{eq:comp-UV}
\end{equation}

\item From equations \eref{eq:comp-QU} and
    \eref{eq:comp-QZfwd}, or equivalently \eref{eq:comp-RU}
    and \eref{eq:comp-RZfwd}, we have
\begin{equation}
  \sum_{x_{i}} {U_{ij,i}}^{(t)}(y_{i,j},x_{i})
  = {Z_{ij}}^{(t+1)}(y_{i,j})
  \qquad \forall i, \forall j \in \partial i
  \, . \label{eq:comp-UZfwd}
\end{equation}

\end{trivlist}
Let us finally remark that the largely redundant form, in which
we have written the compatibility constraints, is not an end in
itself. Indeed, it will enable us to get a very compact proof
of the fact that our method actually produces compatible
solutions. This will be shown, along with the computation
algorithm, in \ref{subapp:algo}.

\subsection{The variational functional}

As mentioned in section~\ref{subsec:specific_approximations},
we derive the PQR free-energy functional (denoted as
$\mathcal{F}_\mathrm{PQR}$) from equation
\eref{eq:free_energy}, replacing the entropic term
$\mathcal{S}$ with $\mathcal{S}_\mathrm{PQR}$ (equation
\eref{eq:entropia-PQR}). Taking into account that we are
interested in solutions respecting the constraint equations,
specifically \eref{eq:comp-PS}, \eref{eq:comp-QT},
\eref{eq:comp-QZ}, \eref{eq:comp-RU}, \eref{eq:comp-RV}, it is
possible to rewrite the functional in such a way that the
subcluster distributions appear only in the logarithmic terms,
that is in the following form
\begin{eqnarray}
  \mathcal{F}_\mathrm{PQR} [ P,Q,R,S,T,U,V,Z ]
  = \nonumber \\ \left.
  = \sum_{t=0}^{\tau-1} \sum_{i} \sum_{y_{i},x_{i,\partial i}} {P_{i}}^{(t)}(y_{i},x_{i,\partial i})
  \ln \frac{
  {P_{i}}^{(t)}(y_{i},x_{i,\partial i})
  }{
  {w_{i}}^{(t)}(y_{i}|x_{i,\partial i})
  {S_{i}}^{(t)}(x_{i,\partial i})
  }
  \right. + \nonumber \\ \left.
  + \sum_{t=0}^{\tau-1} \sum_{ij} \sum_{y_{i,j},x_{i,j}} {Q_{ij}}^{(t)}(y_{i,j},x_{i,j})
  \ln \frac{
  {Q_{ij}}^{(t)}(y_{i,j},x_{i,j})
  {Z_{ij}}^{(t)}(x_{i,j})
  }{
  {T_{i,ij}}^{(t)}(y_{i},x_{i,j})
  {T_{j,ij}}^{(t)}(y_{j},x_{i,j})
  }
  \right. + \nonumber \\ \left.
  + \sum_{t=0}^{\tau-1} \sum_{i} \sum_{y_{i,\partial i},x_{i}} {R_{i}}^{(t)}(y_{i,\partial i},x_{i})
  \ln \frac{
  {R_{i}}^{(t)}(y_{i,\partial i},x_{i})
  {{V_{i}}^{(t)}(y_{i},x_{i})}^{d_{i}-1}
  }{
  \prod_{j \in \partial i}
  {U_{ij,i}}^{(t)}(y_{i,j},x_{i})
  }
  \right. , \label{eq:funz-PQR}
\end{eqnarray}
where the inner sums run over all possible configurations of
P-, Q-, and R-clusters, respectively. Recall that the terms
${w_{i}}^{(t)}(y_{i}|x_{i,\partial i})$ are transition
probabilities, so that by construction we assume they satisfy
the following normalization conditions
\begin{equation}
  \sum_{y_{i}} {w_{i}}^{(t)}(y_{i}|x_{i,\partial i}) = 1
  \qquad \forall i
  \, . \label{eq:norm-W}
\end{equation}
Note also that, given the M\"obius numbers of the PQR
approximation (see table \ref{tab:superposition_coefficients}
and equation \eref{eq:entropia-PQR}), the probability
distributions ${S_{i}}^{(0)}$ (i.e. the S-cluster distributions
at time ${t=0}$) and ${Z_{ij}}^{(0)}$ (i.e. the Z-cluster
distributions at time ${t=0}$) actually do not belong to the
entropic term. We have nonetheless included them in the
free-energy expression, to take into account the part of
energetic term that represents initial conditions, assuming
therefore
\begin{subequations} \label{eq:iniz-SZ}
\begin{eqnarray}
  {S_{i}}^{(0)}(x_{i, \partial i})
  & \equiv {p_{i}}^{(0)}(x_{i}) \prod_{j \in \partial i} {p_{j}}^{(0)}(x_{j})
  & \qquad \forall i
  \, , \label{eq:iniz-S} \\
  {Z_{ij}}^{(0)}(x_{i,j})
  & \equiv {p_{i}}^{(0)}(x_{i}) {p_{j}}^{(0)}(x_{j})
  & \qquad \forall ij
  \, . \label{eq:iniz-Z}
\end{eqnarray}
\end{subequations}

Now, a crucial observation is that the variational functional
in the form \eref{eq:funz-PQR} is made up of a sum of
Kullback-Leibler divergence terms, and therefore that the
absolute minimum conditions (in which every term equals $0$)
are, for ${t=0,1,\dots,\tau-1}$,
\begin{subequations}
\begin{eqnarray}
  {P_{i}}^{(t)}(y_{i},x_{i,\partial i})
  & = {w_{i}}^{(t)}(y_{i}|x_{i,\partial i})
  {S_{i}}^{(t)}(x_{i,\partial i})
  \qquad & \forall i
  \, , \label{eq:algo-P} \\
  {Q_{ij}}^{(t)}(y_{i,j},x_{i,j})
  & = \frac{
  {T_{i,ij}}^{(t)}(y_{i},x_{i,j})
  {T_{j,ij}}^{(t)}(y_{j},x_{i,j})
  }{
  {Z_{ij}}^{(t)}(x_{i,j})
  }
  \qquad & \forall ij
  \, , \label{eq:algo-Q} \\
  {R_{i}}^{(t)}(y_{i,\partial i},x_{i})
  & = \frac{
  \prod_{j \in \partial i} {U_{ij,i}}^{(t)}(y_{i,j},x_{i})
  }{
  {{V_{i}}^{(t)}(y_{i},x_{i})}^{d_{i}-1}
  } \qquad & \forall i
  \, . \label{eq:algo-R}
\end{eqnarray}
\label{eq:algo-PQR}
\end{subequations}
Actually, the minimum condition of a Kullback-Leibler
divergence holds provided its two arguments (that is in our
case the two sides of each equation \eref{eq:algo-PQR}) are
``good'' (i.e. normalized) probability distributions. Assuming
of course the normalization of all cluster distributions, it is
easy to see that the right-hand sides of equations
\eref{eq:algo-P}, \eref{eq:algo-Q} and \eref{eq:algo-R} are
actually good distributions, if, respectively, the
normalization condition \eref{eq:norm-W} and the compatibility
conditions \eref{eq:comp-TZ} and \eref{eq:comp-UV} are
verified.

Let us finally observe that equations \eref{eq:algo-PQR} are
somehow ``weak'', meaning that they do not define a point in
the space of variational parameters, but rather a manifold. In
order to determine completely the variational parameters, that
is the cluster distributions ${P,Q,R,S,T,U,V,Z}$, we have to
intersect the aforementioned manifold with that defined by the
constraints, that is to solve simultaneously the minimum
equations \eref{eq:algo-PQR} with the compatibility equations
\eref{eq:vinc-P}, \eref{eq:vinc-Q}, \eref{eq:vinc-R} (whence
also \eref{eq:comp-SZ}, \eref{eq:comp-TZ}, \eref{eq:comp-TV},
\eref{eq:comp-UV}, \eref{eq:comp-UZfwd}) and the
normalizations. Let us stress the fact that it is not necessary
to resort to the Lagrange multiplier method to solve the
constrained minimum problem, just because it turns out that the
intersection between the two manifolds is nonempty. In general,
the latter is not a necessary condition for the existence of a
constrained minimum. In our specific problem, such a condition
arises from the fact that the exponentials (Boltzmann factors)
of the coupling terms are actually (transition) probabilities,
and therefore that they enjoy the normalization property
\eref{eq:norm-W}.

\subsection{Computation algorithm}

\label{subapp:algo}

In this section we present an iterative procedure, which solves
the problem described above. In particular we see that,
starting with a set of $S$ and $Z$ distributions satisfying
\eref{eq:comp-SZ} at a given timestep $t$ (initially ${t=0}$,
with the distributions defined by \eref{eq:iniz-SZ}), one can
determine all the other distributions $P,Q,R$ and $T,U,V$ at
timestep $t$, as well as $S$ and $Z$ at timestep ${t+1}$,
satisfying all the required equations at time $t$ and
\eref{eq:comp-SZ} at time $t+1$. Moreover, all the computed
distributions turn out to be correctly normalized.\footnote{In
the numerical implementation, it turns out to be necessary to
restore normalization explicitly.} The steps of the procedure
are described in detail below, and summarized as a flow chart
in figure~\ref{fig:flow-chart}.

\setlength{\unitlength}{1mm}

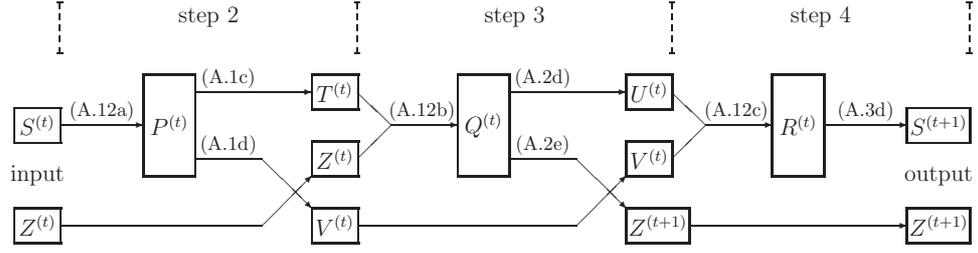
\begin{figure}
\flushright
\resizebox{132mm}{!}{
  \begin{picture}(176,50)(-3,-25)
    %\put(-3,-25){\framebox(176,50){}}

    \thicklines
    \put(0,-3){\framebox(8,6){${S}^{(t)}$}}
    \thinlines
    \put(8,0){\vector(1,0){15}}
    \put(8,0){\makebox(15,5){\footnotesize \eref{eq:algo-P}}}
    \put(0,-11.5){\makebox(8,5){input}}

    \thicklines
    \put(23,-9){\framebox(9,18){${P}^{(t)}$}}
    \thinlines
    \put(32,6){\vector(1,0){21}}
    \put(32,6){\makebox(12,5){\footnotesize \eref{eq:comp-PT}}}
    \thicklines
    \put(53,3){\framebox(8,6){${T}^{(t)}$}}
    \put(53,-9){\framebox(8,6){${Z}^{(t)}$}}
    \put(53,-21){\framebox(8,6){${V}^{(t)}$}}
    \thinlines
    \put(32,-6){\line(1,0){12}}
    \put(44,-6){\vector(1,-1){9}}
    \put(32,-6){\makebox(12,5){\footnotesize \eref{eq:comp-PV}}}
    \put(44,-18){\vector(1,1){9}}
    \put(8,-18){\line(1,0){36}}
    \thicklines
    \put(0,-21){\framebox(8,6){${Z}^{(t)}$}}
    \thinlines
    \put(61,6){\line(1,-1){6}}
    \put(61,-6){\line(1,1){6}}
    \put(67,0){\vector(1,0){12}}
    \put(66.5,0){\makebox(12,5){\footnotesize \eref{eq:algo-Q}}}

    \thicklines
    \put(79,-9){\framebox(9,18){${Q}^{(t)}$}}
    \thinlines
    \put(88,6){\vector(1,0){21}}
    \put(88,6){\makebox(12,5){\footnotesize \eref{eq:comp-QU}}}
    \thicklines
    \put(109,3){\framebox(8,6){${U}^{(t)}$}}
    \put(109,-9){\framebox(8,6){${V}^{(t)}$}}
    \put(109,-21){\framebox(11,6){${Z}^{(t+1)}$}}
    \thinlines
    \put(88,-6){\line(1,0){12}}
    \put(100,-6){\vector(1,-1){9}}
    \put(88,-6){\makebox(12,5){\footnotesize \eref{eq:comp-QZfwd}}}
    \put(100,-18){\vector(1,1){9}}
    \put(61,-18){\line(1,0){39}}
    \put(117,6){\line(1,-1){6}}
    \put(117,-6){\line(1,1){6}}
    \put(123,0){\vector(1,0){12}}
    \put(122.5,0){\makebox(12,5){\footnotesize \eref{eq:algo-R}}}

    \thicklines
    \put(135,-9){\framebox(9,18){${R}^{(t)}$}}
    \thinlines
    \put(144,0){\vector(1,0){15}}
    \put(144,0){\makebox(15,5){\footnotesize \eref{eq:comp-RSfwd}}}
    \thicklines
    \put(159,-3){\framebox(11,6){${S}^{(t+1)}$}}
    \put(159,-21){\framebox(11,6){${Z}^{(t+1)}$}}
    \thinlines
    \put(120,-18){\vector(1,0){39}}
    \put(159,-11.5){\makebox(11,5){output}}

    \put(8,13){\dashbox(0,9){}}
    \put(8,17){\makebox(53,5){step 2}}
    \put(61,13){\dashbox(0,9){}}
    \put(61,17){\makebox(56,5){step 3}}
    \put(117,13){\dashbox(0,9){}}
    \put(117,17){\makebox(53,5){step 4}}
    \put(170,13){\dashbox(0,9){}}

  \end{picture}
}
  \caption
  {
    Flow chart of the computational procedure (one timestep) for the PQR
    approximation. Large and small boxes represent probability distributions
    for, respectively, maximal-clusters and subclusters.
    Arrows going into maximal-cluster boxes represent the minimum
    equations \eref{eq:algo-PQR}, whereas arrows coming out of
    such boxes represent marginalization equations.
  }
  \label{fig:flow-chart}
\end{figure}

\begin{enumerate}

\item Let us begin, at a given timestep $t$ (initially
    ${t=0}$) with a set of (normalized) S-cluster
    distributions ${S_{i}}^{(t)}(x_{i,\partial i})$
    ($\forall i$) and Z-cluster distributions
    ${Z_{ij}}^{(t)}(x_{i,j})$ ($\forall ij$), that we
    assume to be compatible, i.e. satisfying equations
    \eref{eq:comp-SZ} (which is obviously verified by the
    initial distributions \eref{eq:iniz-SZ}).

\item \label{item:step2} We determine the P-cluster
    distributions at time $t$ by means of equations
    \eref{eq:algo-P}. Thanks to \eref{eq:norm-W}, we see
    that such distributions satisfy equations
    \eref{eq:comp-PS}. Moreover, since \eref{eq:comp-PS}
    and \eref{eq:comp-SZ} hold simultaneously, also
    \eref{eq:comp-PZ} turns out to be satisfied. We then
    determine the following marginals.
    \begin{itemize}

    \item[--] T-cluster distributions
        ${T_{i,ij}}^{(t)}(y_{i},x_{i,j})$ ($\forall i,
        \forall j \in \partial i$) by equations
        \eref{eq:comp-PT}. Since \eref{eq:comp-PT} and
        \eref{eq:comp-PZ} hold simultaneously, such
        distributions also satisfy \eref{eq:comp-TZ}.

   \item[--] V-cluster distributions
       ${V_{i}}^{(t)}(y_{i},x_{i})$ ($\forall i$) by
       equations \eref{eq:comp-PV}. Since
       \eref{eq:comp-PV} and \eref{eq:comp-PT} hold
       simultaneously, also \eref{eq:comp-TV} turns out
       to be satisfied.

   \end{itemize}

\item \label{item:step3} We determine the Q-cluster
    distributions at time $t$ by means of equations
    \eref{eq:algo-Q}. Thanks to \eref{eq:comp-TZ}, we see
    that such distributions satisfy equations
    \eref{eq:comp-QT} and \eref{eq:comp-QZ}. Moreover,
    since \eref{eq:comp-QT} and \eref{eq:comp-TV} hold
    simultaneously, also \eref{eq:comp-QV} turns out to be
    satisfied. We then determine the following marginals.
    \begin{itemize}

    \item[--] U-cluster distributions
        ${U_{ij,i}}^{(t)}(y_{i,j},x_{i})$ ($\forall i,
        \forall j \in \partial i$) by equations
        \eref{eq:comp-QU}. Since \eref{eq:comp-QU} and
        \eref{eq:comp-QV} hold simultaneously, such
        distributions also satisfy \eref{eq:comp-UV}.

    \item[--] Z-cluster distributions (at the next
        timestep) ${Z_{ij}}^{(t+1)}(y_{i,j})$ ($\forall
        ij$) by equations \eref{eq:comp-QZfwd}. Since
        \eref{eq:comp-QZfwd} and \eref{eq:comp-QU} hold
        simultaneously, also \eref{eq:comp-UZfwd} turns
        out to be satisfied.

    \end{itemize}

\item \label{item:step4} We determine the R-cluster
    distributions at time $t$ by means of equations
    \eref{eq:algo-R}. Thanks to \eref{eq:comp-UV}, we see
    that such distributions satisfy equations
    \eref{eq:comp-RU} and \eref{eq:comp-RV}. Moreover,
    since \eref{eq:comp-RU} and \eref{eq:comp-UZfwd} hold
    simultaneously, also \eref{eq:comp-RZfwd} turns out to
    be satisfied. We then determine the following
    marginals.

    \begin{itemize}

    \item[--] S-cluster distributions (at the next
        timestep) ${S_{i}}^{(t+1)}(y_{i,\partial i})$
        ($\forall i$) by equations
        \eref{eq:comp-RSfwd}. Since
        \eref{eq:comp-RSfwd} and \eref{eq:comp-RZfwd}
        hold simultaneously, such distributions also
        satisfy \eref{eq:comp-SZ} at time ${t+1}$.

    \end{itemize}

\item We repeat the procedure from step~\eref{item:step2}
    for the distributions at time ${t+1}$.

\end{enumerate}
Let us stress the fact that, in the above procedure, all the
compatibility conditions stated in \ref{subapp:comp} have been
either used explicitly to compute the marginals or proved to be
satisfied by the computed distributions.

\subsection{The \emph{diamond} approximation}

\label{subapp:diamond}

In this section we show that, as mentioned in
section~\ref{subsec:related_works}, in the case of a pure
neighbor-conditioned dynamics the PQR approximation coincides
with the diamond approximation, previously proposed by one of
us in reference~\cite{Pelizzola2013}.

First of all it is useful to define some auxiliary clusters,
namely, the B-clusters, made up of two neighbor sites at
subsequent timesteps $t$ and ${t+1}$, for
${t=0,1,\dots,\tau-1}$, and the C-clusters, made up of all the
neighbors of a given site at a given timestep $t$, for
${t=0,1,\dots,\tau}$. The related distributions are denoted as
\begin{eqnarray}
  {B_{i,j}}^{(t)}(y_{i},x_{j})
  & \triangleq
  \mathbb{P} \{ {\xi_{i}}^{(t+1)} = y_{i} ,\, {\xi_{j}}^{(t)} = x_{j}\}
  & \qquad \forall i, \forall j \in \partial i
  \, , \\
  {C_{i}}^{(t)}(x_{\partial i})
  & \triangleq
  \mathbb{P} \{ {\xi_{\partial i}}^{(t)} = x_{\partial i} \}
  & \qquad \forall i
  \, .
\end{eqnarray}
These distributions can be written as marginals of maximal
cluster distributions, so that they inherit all the required
compatibility conditions. We shall use in particular the
following relationship
\begin{equation}
  {B_{i,j}}^{(t)}(y_{i},x_{j})
  = \sum_{x_{i, \partial i \setminus j}} {P_{i}}^{(t)}(y_{i},x_{i,\partial i})
  \qquad \forall i, \forall j \in \partial i
  \, , \label{eq:comp-PB}
\end{equation}
which, together with
\begin{equation}
  {A_{i}}^{(t+1)}(y_{i})
  = \sum_{x_{i, \partial i}} {P_{i}}^{(t)}(y_{i},x_{i,\partial i})
  \qquad \forall i
  \, , \label{eq:comp-PAfwd}
\end{equation}
also imply
\begin{equation}
  {A_{i}}^{(t+1)}(y_{i})
  = \sum_{x_{j}} {B_{i,j}}^{(t)}(y_{i},x_{j})
  \qquad \forall i, \forall j \in \partial i
  \, . \label{eq:comp-BAfwd}
\end{equation}

Let us now hypothesize that the $S$ and $Z$ distributions at a
given timestep $t$ factor in the following fashion
\begin{subequations} \label{eq:fatt-ipo}
\begin{eqnarray}
  {S_{i}}^{(t)}(x_{i,\partial i})
  & = {A_{i}}^{(t)}(x_{i})
  {C_{i}}^{(t)}(x_{\partial i})
  & \qquad \forall i
  \, , \label{eq:fatt-SAC} \\
  {Z_{ij}}^{(t)}(x_{i,j})
  & = {A_{i}}^{(t)}(x_{i})
  {A_{j}}^{(t)}(x_{j})
  & \qquad \forall ij
  \, , \label{eq:fatt-ZAA}
\end{eqnarray}
\end{subequations}
obviously assuming also that $C$ and $A$ distributions are
compatible, that is they satisfy
\begin{equation}
  \sum_{x_{\partial i \setminus j}} {C_{i}}^{(t)}(x_{\partial i}) =
  {A_{j}}^{(t)}(x_{j})
  \qquad \forall i, \forall j \in \partial i
  \, .
  \label{eq:comp-CA}
\end{equation}
Retracing the steps of the algorithm described in the previous
section, let us show that the hypothesized factorizations are
stable along the time evolution (i.e., our hypotheses imply the
same factorizations at timestep $t+1$), and the evolution
equations are identical to those of
reference~\cite{Pelizzola2013}.

\begin{enumerate}

\item We easily see that the hypotheses \eref{eq:fatt-ipo}
    and \eref{eq:comp-CA} satisfy the compatibility
    \eref{eq:comp-SZ} between $S$ and $Z$ distributions at
    time $t$.

\item Equation \eref{eq:algo-P} with the factorization
    \eref{eq:fatt-SAC} reads
\begin{equation}
  {P_{i}}^{(t)}(y_{i},x_{i,\partial i}) =
  {w_{i}}^{(t)}(y_{i}|x_{\partial i})
  {A_{i}}^{(t)}(x_{i})
  {C_{i}}^{(t)}(x_{\partial i})
  \qquad \forall i
  \, . \label{eq:fatt-PwAC}
\end{equation}
Therefore, equation \eref{eq:comp-PT} together with
\eref{eq:comp-PB} gives
\begin{equation}
  {T_{i,ij}}^{(t)}(y_{i},x_{i,j})
  = {A_{i}}^{(t)}(x_{i}) {B_{i,j}}^{(t)}(y_{i},x_{j})
  \qquad \forall i, \forall j \in \partial i
  \, ,
  \label{eq:fatt-TAB}
\end{equation}
where
\begin{equation}
  {B_{i,j}}^{(t)}(y_{i},x_{j})
  = \sum_{x_{\partial i \setminus j}}
  {w_{i}}^{(t)}(y_{i}|x_{\partial i})
  {C_{i}}^{(t)}(x_{\partial i})
  \qquad \forall i, \forall j \in \partial i
  \, ,
  \label{eq:algo-B}
\end{equation}
whereas equation \eref{eq:comp-PV} together with
\eref{eq:comp-PAfwd} gives
\begin{equation}
  {V_{i}}^{(t)}(y_{i},x_{i})
  = {A_{i}}^{(t+1)}(y_{i}) {A_{i}}^{(t)}(x_{i})
  \qquad \forall i
  \, ,
  \label{eq:fatt-VAA}
\end{equation}
where
\begin{equation}
  {A_{i}}^{(t+1)}(y_{i})
  = \sum_{x_{\partial i}}
  {w_{i}}^{(t)}(y_{i}|x_{\partial i})
  {C_{i}}^{(t)}(x_{\partial i})
  \qquad \forall i
  \, .
  \label{eq:algo-A}
\end{equation}

\item Equation \eref{eq:algo-Q} with the factorizations
    \eref{eq:fatt-TAB} and \eref{eq:fatt-ZAA} reads
\begin{equation}
  {Q_{ij}}^{(t)}(y_{i,j},x_{i,j})
  = {B_{i,j}}^{(t)}(y_{i},x_{j}) {B_{j,i}}^{(t)}(y_{j},x_{i})
  \qquad \forall ij
  \, ,
\end{equation}
Therefore, equation \eref{eq:comp-QU} together with
\eref{eq:comp-BAfwd} gives
\begin{equation}
  {U_{ij,i}}^{(t)}(y_{i,j},x_{i})
  = {A_{i}}^{(t+1)}(y_{i}) {B_{j,i}}^{(t)}(y_{j},x_{i})
  \qquad \forall i, \forall j \in \partial i
  \, ,
  \label{eq:fatt-UAB}
\end{equation}
whereas equation \eref{eq:comp-QZfwd} together with
\eref{eq:comp-BAfwd} gives
\begin{equation}
  {Z_{ij}}^{(t+1)}(y_{i,j})
  = {A_{i}}^{(t+1)}(y_{i}) {A_{j}}^{(t+1)}(y_{j})
  \qquad \forall ij
  \, .
\end{equation}
We thus argue that the $Z$ distributions at time ${t+1}$
take a factorized form analogous to \eref{eq:fatt-ZAA}.

\item Equation \eref{eq:algo-R} with the factorizations
    \eref{eq:fatt-UAB} and \eref{eq:fatt-VAA} reads
\begin{equation}
  {R_{i}}^{(t)}(y_{i,\partial i},x_{i}) =
  {A_{i}}^{(t+1)}(y_{i}) \frac{
  \prod_{j \in \partial i} {B_{j,i}}^{(t)}(y_{j},x_{i})
  }{
  {{A_{i}}^{(t)}(x_{i})}^{d_{i}-1}
  }
  \qquad \forall i
  \, . \label{eq:fatt-RABA}
\end{equation}
Therefore, equation \eref{eq:comp-RSfwd} proves that the
$S$ distributions at time ${t+1}$ take a factorized form
analogous to \eref{eq:fatt-SAC}, namely
\begin{equation}
  {S_{i}}^{(t+1)}(y_{i,\partial i})
  = {A_{i}}^{(t+1)}(y_{i})
  {C_{i}}^{(t+1)}(y_{\partial i})
  \qquad \forall i
  \, ,
\end{equation}
where
\begin{equation}
  {C_{i}}^{(t+1)}(y_{\partial i}) =
  \sum_{x_{i}} \frac{
  \prod_{j \in \partial i} {B_{j,i}}^{(t)}(y_{j},x_{i})
  }{
  {{A_{i}}^{(t)}(x_{i})}^{d_{i}-1}
  }
  \qquad \forall i
  \, . \label{eq:algo-C}
\end{equation}

\end{enumerate}
Note that the ``diamond clusters'' of
reference~\cite{Pelizzola2013} never appear explicitly in the
above discussion. Nevertheless, we can see that, plugging
equation \eref{eq:algo-C} into \eref{eq:algo-B} and
\eref{eq:algo-A}, one obtains evolution equations for B- and
A-cluster distributions, which exactly coincide with those one
could obtain, in reference \cite{Pelizzola2013}, replacing the
diamond cluster distribution (26) into the marginalization
equations (25) and (23).

\section{Other approximations}

\label{app:other_approximations}

In this appendix we discuss the other approximations we have
introduced in the text, namely the PQ and P approximations, and
the M approximation. We do not go into the same level of
detail, as done for the PQR approximation, because the
derivations are in fact very much analogous. We only give the
final results in terms of the respective computation
procedures.

\subsection{PQ and P approximations}

\label{subapp:PQandP}

The PQ approximation can be regarded as a simplification of the
PQR approximation, in which the S-cluster distributions are
factorized in terms of Z-cluster (nearest-neighbor pair) and
A-cluster (single site) distributions, according to
\begin{equation}
  {S_{i}}^{(t)}(x_{i,\partial i}) =
  \frac{
  \prod_{j \in \partial i} {Z_{ij}}^{(t)}(x_{i,j})
  }{
  {{A_{i}}^{(t)}(x_{i})}^{d_{i}-1}
  }
  \qquad \forall i
  \, . \label{eq:fatt-SZA}
\end{equation}
As a consequence, one can replace step~\eref{item:step4} of the
PQR procedure with the previous equation (evaluated at timestep
${t+1}$), where the Z-cluster distributions have been
determined at step~\eref{item:step3}, and the A-cluster
distributions, compatible with the former, i.e. satisfying
\begin{equation}
  \sum_{x_{j} }{Z_{ij}}^{(t)}(x_{i,j})
  = {A_{i}}^{(t)}(x_{i})
  \qquad \forall i, \forall j \in \partial i
  \, , \label{eq:comp-ZA}
\end{equation}
can be written by the obvious marginalization
\eref{eq:comp-PAfwd}. In conclusion, as previously mentioned,
the PQ approximation turns out to be equivalent to the pairwise
approximation, considered by different
authors~\cite{WangTangStanleyBraunstein2016,PastorsatorrasCastellanoVanmieghemVespignani2015,PetermannDelosrios2004}.
Such an equivalence might indeed be not fully evident, because
in the cited papers the approximations are immediately
specialized to a particular model (namely, the SIS model) and
to the continuous-time limit, whereas our treatment is more
general, at least under the assumptions of a discrete-time
dynamics and a transition probability factorized according to
\eref{eq:transition_probabilities}.

The P approximation can be regarded as a further
simplification, in which also the pair distributions are
factorized, as in equation \eref{eq:fatt-ZAA}, so that the
S-cluster distributions \eref{eq:fatt-SZA} read
\begin{equation}
  {S_{i}}^{(t)}(x_{i,\partial i}) =
  {A_{i}}^{(t)}(x_{i})
  \prod_{j \in \partial i} {A_{j}}^{(t)}(x_{j})
  \qquad \forall i
  \, . \label{eq:fatt-SAA}
\end{equation}
As a consequence, one can skip also step~\eref{item:step3} of
the PQR procedure, because there is no need to compute pair
distributions at the next timestep, to be able to close the
equations. Indeed, replacing equation \eref{eq:fatt-SAA} into
\eref{eq:algo-P} and then into \eref{eq:comp-PAfwd}, one
obtains a unique time-evolution equation for the single-site
distributions, namely
\begin{equation}
  {A_{i}}^{(t+1)}(y_{i})
  = \sum_{x_{i,\partial i}}
  {w_{i}}^{(t)}(y_{i}|x_{i,\partial i})
  {A_{i}}^{(t)}(x_{i})
  \prod_{j \in \partial i} {A_{j}}^{(t)}(x_{j})
  \qquad \forall i
  \, ,
\end{equation}
which coincides with the (quenched) mean-field
approximation~\cite{WangTangStanleyBraunstein2016,PastorsatorrasCastellanoVanmieghemVespignani2015,PetermannDelosrios2004}.

This argument also proves that, for a pure neighbor-conditioned
dynamics, for which \eref{eq:fatt-ZAA} holds, the PQ
approximation degenerates into the P approximation, and
therefore that both coincide with the \emph{star} approximation
of reference~\cite{Pelizzola2013}, as previously stated in
section~\ref{subsec:related_works}.

\subsection{M approximation}

\label{subapp:M}

The iterative procedure for the M approximation turns out to
coincide with that of the PQR approximation until
step~\eref{item:step3}. Step~\eref{item:step4} is modified in
that the R-cluster distributions are no longer computed by
equation \eref{eq:algo-R}, but rather as marginals of the
M-cluster distributions, namely
\begin{equation}
  {R_{i}}^{(t)}(y_{i,\partial i},x_{i}) =
  \sum_{x_{\partial i}}
  {M_{i}}^{(t)}(y_{i,\partial i},x_{i,\partial i})
  \qquad \forall i
  \, , \label{eq:comp-MR}
\end{equation}
the latter being previously computed as
\begin{equation}
  {M_{i}}^{(t)}(y_{i,\partial i},x_{i,\partial i}) =
  {P_{i}}^{(t)}(y_{i},x_{i,\partial i})
  \prod_{j \in \partial i} \frac{
  {T_{j,ij}}^{(t)}(y_{j},x_{i,j})
  }{
  {Z_{ij}}^{(t)}(x_{i,j})
  }
  \qquad \forall i
  \, . \label{eq:algo-M}
\end{equation}

In the simpler case of a pure neighbor-conditioned dynamics, we
have shown for the PQR approximation that the factorizations
\eref{eq:fatt-ZAA}, \eref{eq:fatt-PwAC}, and \eref{eq:fatt-TAB}
hold. These equations remain valid also for the M
approximation, because they stay within step~\eref{item:step3}
of the procedure, so that we can plug them into
\eref{eq:algo-M} and then into \eref{eq:comp-MR}. Taking into
account also equation \eref{eq:algo-A} (still valid), we obtain
exactly equation \eref{eq:fatt-RABA}. This argument proves
that, for a pure neighbor-conditioned dynamics, the M
approximation degenerates into the PQR approximation, and
therefore that both coincide with the \emph{diamond}
approximation of reference~\cite{Pelizzola2013}, as previously
stated in section~\ref{subsec:related_works}.

\section{Continuous-time limit}

\label{app:continuous_time}

The approximation schemes we have presented in this paper can
be used as well to analyze continuous-time processes (in a
discretized form), or equivalently discrete-time processes with
sequential dynamics. This can be done by defining the following
transition probabilities
\begin{equation}
  {w_{i}}^{(t)}(y_{i}|x_{i,\partial i})
  = (1 - \tau) \delta(y_{i},x_{i})
  + \tau \tilde{w}{\vphantom{w}_{i}}^{(t)}(y_{i}|x_{i,\partial i})
  \, , \label{eq:transition_probabilities_continuous_time}
\end{equation}
where it is understood that ${\tau \to 0}$, whereas
$\delta(\cdot,\cdot)$ denotes a Kronecker delta and
$\tilde{w}{\vphantom{w}_{i}}^{(t)}(y_{i}|x_{i,\partial i})$ are
suitable conditional probabilities, normalized in the proper
way
\begin{equation}
  \sum_{y_{i}} \tilde{w}{\vphantom{w}_{i}}^{(t)}(y_{i}|x_{i,\partial i}) = 1
  \, .
\end{equation}
Equation \eref{eq:transition_probabilities_continuous_time} can
be interpreted as follows. With a large probability (${1 -
\tau}$) each vertex variable $\xi_{i}^{(t)}$ stays in its
current value, and with a small probability ($\tau$) it is
allowed to undergo a transition, governed by the probabilities
$\tilde{w}{\vphantom{w}_{i}}^{(t)}$. For the case of sequential
dynamics, ${\tau = 1/N}$ is the probability of choosing one
specific vertex out of $N$. Alternatively, for the case of
continuous-time processes, one can think of $\tau$ as a small
time interval, and of
$\tilde{w}{\vphantom{w}_{i}}^{(t)}(y_{i}|x_{i,\partial i})$ as
the \emph{rate} of the transition ${x_{i} \to y_{i}}$ (for
$y_{i} \neq x_{i}$), which is assumed to depend on the
neighborhood configuration $x_{\partial i}$.

As mentioned in the text, we can show that, in the limit ${\tau
\to 0}$ (continuous-time limit), the PQR and PQ approximations
become equivalent. Making use of equations \eref{eq:algo-P} and
\eref{eq:comp-PT} (step~\eref{item:step2} of the computation
algorithm) with the transition probabilities
\eref{eq:transition_probabilities_continuous_time}, and taking
into account the compatibility equations \eref{eq:comp-SZ}, the
T-cluster distributions can be written as
\begin{subequations}
\begin{eqnarray}
  {T_{i,ij}}^{(t)}(y_{i},x_{i,j})
  & = \sum_{x_{\partial i \setminus j}}
  {w_{i}}^{(t)}(y_{i}|x_{i,\partial i})
  {S_{i}}^{(t)}(x_{i,\partial i})
  = \label{eq:ct-T} \\
  & = (1 - \tau) \delta(y_{i},x_{i}) {Z_{ij}}^{(t)}(x_{i,j})
  + \tau \tilde{T}{\vphantom{T}_{i,ij}}^{(t)}(y_{i},x_{i,j})
  \, , \label{eq:ct-T_espansa}
\end{eqnarray}
\end{subequations}
where we have defined\footnote{Comparing equations
\eref{eq:ct-T} and \eref{eq:ct-Ttilde}, one can argue that
$\tilde{T}$ is the T-cluster distribution one would get if the
transition probabilities were barely $\tilde{w}$. In the
following we use the same notation as well for other cluster
types.}
\begin{equation}
  \tilde{T}{\vphantom{T}_{i,ij}}^{(t)}(y_{i},x_{i,j})
  \triangleq \sum_{x_{\partial i \setminus j}}
  \tilde{w}{\vphantom{w}_{i}}^{(t)}(y_{i}|x_{i,\partial i})
  {S_{i}}^{(t)}(x_{i,\partial i})
  \, . \label{eq:ct-Ttilde}
\end{equation}
Making use of equations \eref{eq:algo-Q} and
\eref{eq:comp-QZfwd} (step~\eref{item:step3} of the computation
algorithm), still with the transition probabilities
\eref{eq:transition_probabilities_continuous_time}, the
Z-cluster distributions at the next timestep (now denoted as
${t+\tau}$ rather than ${t+1}$) can be written as
\begin{equation}
  \fl
  {Z_{ij}}^{(t+\tau)}(y_{i,j})
  = (1-2\tau) {Z_{ij}}^{(t)}(y_{i,j})
  + \tau \left[ \tilde{B}{\vphantom{B}_{i,j}}^{(t)}(y_{i},y_{j})
  + \tilde{B}{\vphantom{B}_{j,i}}^{(t)}(y_{j},y_{i}) \right]
  + \mathcal{O}(\tau^2)
  \, , \label{eq:ct-Zfwd}
\end{equation}
where we have defined
\begin{equation}
  \tilde{B}{\vphantom{B}_{i,j}}^{(t)}(y_{i},x_{j})
  \triangleq \sum_{x_{i, \partial i \setminus j}}
  \tilde{w}{\vphantom{w}_{i}}^{(t)}(y_{i}|x_{i,\partial i})
  {S_{i}}^{(t)}(x_{i,\partial i})
  \, . \label{eq:ct-Btilde}
\end{equation}
Moreover, by \eref{eq:algo-P} and \eref{eq:comp-PAfwd} we get
\begin{equation}
  {A_{i}}^{(t+\tau)}(y_{i})
  = (1-\tau) {A_{i}}^{(t)}(y_{i})
  + \tau \tilde{A}{\vphantom{A}_{i}}^{(t+\tau)}(y_{i})
  \, , \label{eq:ct-Afwd}
\end{equation}
where
\begin{equation}
  \tilde{A}{\vphantom{A}_{i}}^{(t+\tau)}(y_{i})
  \triangleq \sum_{x_{i, \partial i}}
  \tilde{w}{\vphantom{w}_{i}}^{(t)}(y_{i}|x_{i,\partial i})
  {S_{i}}^{(t)}(x_{i,\partial i})
  \, . \label{eq:ct-Atildefwd}
\end{equation}
In principle, taking the limit ${\tau \to 0}$ of equations
\eref{eq:ct-Zfwd} and \eref{eq:ct-Afwd}, one can obtain a
system of ordinary differential equations for the pair and site
distributions. Indeed, equations \eref{eq:ct-Zfwd} and
\eref{eq:ct-Afwd} can be viewed as discretized forms thereof,
which are directly suitable for numerical implementation.

Let us observe that equations \eref{eq:ct-Zfwd} and
\eref{eq:ct-Afwd} hold for both the PQ and PQR approximations,
since steps \eref{item:step2} and~\eref{item:step3} of the
computational procedure are identical, as mentioned in
\ref{subapp:PQandP}. For the PQ approximation, these equations
are immediately closed, because the $S$ distribution (appearing
in the expressions of $\tilde{A}$ and $\tilde{B}$) directly
depends on $Z$ and $A$, according to equation
\eref{eq:fatt-SZA}. As we shall see below, the reason why the
PQ and PQR approximations turn out to be equivalent is that the
difference between the $S$ distributions for the two cases is
$\mathcal{O}(\tau)$ for ${\tau \to 0}$, so that, being
multiplied by $\tau$ in both equations \eref{eq:ct-Zfwd} and
\eref{eq:ct-Afwd}, its contribution becomes irrelevant. Let us
first consider the PQ approximation. Equation
\eref{eq:fatt-SZA} at time ${t+\tau}$ reads
\begin{equation}
  {S_{i}}^{(t+\tau)}(y_{i,\partial i})
  = \frac{
  \prod_{j \in \partial i} {Z_{ij}}^{(t+\tau)}(y_{i,j})
  }{
  {{A_{i}}^{(t+\tau)}(y_{i})}^{d_{i}-1}
  }
  \, . \label{eq:ct-fatt-SZA}
\end{equation}
Plugging equations \eref{eq:ct-Zfwd} and \eref{eq:ct-Afwd} into
the latter, we easily get
\begin{equation}
  {S_{i}}^{(t+\tau)}(y_{i,\partial i})
  = \frac{
  \prod_{j \in \partial i} {Z_{ij}}^{(t)}(y_{i,j})
  }{
  {{A_{i}}^{(t)}(y_{i})}^{d_{i}-1}
  }
  + \mathcal{O}(\tau)
  \, . \label{eq:ct-Sfwd}
\end{equation}
As far as the PQR approximation is concerned, the $S$
distribution is computed by equation \eref{eq:comp-RSfwd} as a
marginal of the $R$ distribution, which in turn depends on the
$U$ and $V$ distributions via equation \eref{eq:algo-R}
(step~\eref{item:step4} of the numerical procedure). Using
equations \eref{eq:comp-QU} and \eref{eq:algo-Q}
(step~\eref{item:step3} of the procedure) with the $T$
distribution expressed by \eref{eq:ct-T_espansa}, we can write
\begin{equation}
  {U_{ij,i}}^{(t)}(y_{i,j},x_{i})
  = \delta(y_{i},x_{i}) {Z_{ij}}^{(t)}(y_{i,j})
  + \mathcal{O}(\tau)
  \, , \label{eq:ct-U}
\end{equation}
hence, from equations \eref{eq:comp-UV} and \eref{eq:comp-ZA},
\begin{equation}
  {V_{i}}^{(t)}(y_{i},x_{i})
  = \delta(y_{i},x_{i}) {A_{i}}^{(t)}(y_{i})
  + \mathcal{O}(\tau)
  \, . \label{eq:ct-V}
\end{equation}
Plugging equations \eref{eq:ct-U} and \eref{eq:ct-V} into
\eref{eq:algo-R}, we get
\begin{equation}
  {R_{i}}^{(t)}(y_{i,\partial i},x_{i})
  = \delta(y_{i},x_{i}) \frac{
  \prod_{j \in \partial i} {Z_{ij}}^{(t)}(y_{i,j})
  }{
  {{A_{i}}^{(t)}(y_{i})}^{d_{i}-1}
  }
  + \mathcal{O}(\tau)
  \, .
\end{equation}
The latter expression, marginalized via \eref{eq:comp-RSfwd},
yields again \eref{eq:ct-Sfwd}, which is equivalent to the PQ
result.

\ack We warmly express our thanks to Gino Del Ferraro, Eduardo
Dominguez and Federico Ricci-Tersenghi for useful discussions
on the subject.

\section*{References}

\bibliography{bibliografia}

\begin{thebibliography}{10}

\bibitem{Kikuchi1951}
R.~Kikuchi.
\newblock A theory of cooperative phenomena.
\newblock {\em Phys. Rev.}, 81:988, 1951.

\bibitem{PlischkeBergersen1994}
M.~Plischke and B.~Bergersen.
\newblock {\em Equilibrium statistical physics}.
\newblock World Scientific Publishing, Singapore, 1994.

\bibitem{Pelizzola2005}
A.~Pelizzola.
\newblock Cluster variation method in statistical physics and probabilistic
  graphical models.
\newblock {\em J. Phys. A: Math. Gen.}, 38:R309, 2005.

\bibitem{YedidiaFreemanWeiss2005}
J.~S. Yedidia, W.~T. Freeman, and Y.~Weiss.
\newblock Constructing free-energy approximations and generalized belief
  propagation algorithms.
\newblock {\em IEEE Trans. Inform. Theory}, 51:2282, 2005.

\bibitem{WangTangStanleyBraunstein2016}
W.~{Wang}, M.~{Tang}, H.~E. {Stanley}, and L.~A. {Braunstein}.
\newblock Unification of theoretical approaches for epidemic spreading on
  complex networks.
\newblock {\em Rep. Prog. Phys.}, 80:036603, 2017.

\bibitem{PastorsatorrasCastellanoVanmieghemVespignani2015}
R.~Pastor-Satorras, C.~Castellano, P.~Van~Mieghem, and A.~Vespignani.
\newblock Epidemic processes in complex networks.
\newblock {\em Rev. Mod. Phys.}, 87:925, 2015.

\bibitem{PetermannDelosrios2004}
T.~Petermann and P.~De~Los~Rios.
\newblock Cluster approximations for epidemic processes: a systematic
  description of correlations beyond the pair level.
\newblock {\em J. Theor. Biol.}, 229:1, 2004.

\bibitem{CrisantiSompolinsky1988}
A.~Crisanti and H.~Sompolinsky.
\newblock Dynamics of spin systems with randomly asymmetric bonds:
  \protect{Ising} spins and \protect{Glauber} dynamics.
\newblock {\em Phys. Rev. A}, 37:4865, 1988.

\bibitem{RoudiAurellHertz2009}
Y.~Roudi, E.~Aurell, and J.~A. Hertz.
\newblock Statistical physics of pairwise probability models.
\newblock {\em Front. Comput. Neurosci.}, 3:22, 2009.

\bibitem{AldanaCoppersmithKadanoff2003}
M.~Aldana, S.~Coppersmith, and L.~P. Kadanoff.
\newblock Boolean dynamics with random couplings.
\newblock In {\em Perspectives and Problems in Nolinear Science}, page~23.
  Springer, 2003.

\bibitem{CastellanoFortunatoLoreto2009}
C.~Castellano, S.~Fortunato, and V.~Loreto.
\newblock Statistical physics of social dynamics.
\newblock {\em Rev. Mod. Phys.}, 81:591, 2009.

\bibitem{SchweitzerBehera2015}
F.~Schweitzer and L.~Behera.
\newblock Neighborhood approximations for non-linear voter models.
\newblock {\em Entropy}, 15:7658, 2015.

\bibitem{SucheckiEguiluzSanmiguel2005}
K.~Suchecki, V.~M. Egu\'{\i}luz, and M.~San~Miguel.
\newblock Voter model dynamics in complex networks: Role of dimensionality,
  disorder, and degree distribution.
\newblock {\em Phys. Rev. E}, 72:036132, 2005.

\bibitem{ChouMallickZia2011}
T.~Chou, K.~Mallick, and R.~K.~P. Zia.
\newblock Non-equilibrium statistical mechanics: from a paradigmatic model to
  biological transport.
\newblock {\em Rep. Prog. Phys.}, 74:116601, 2011.

\bibitem{BarratBarthelemyVespignani2008}
A.~Barrat, M.~Barthelemy, and A.~Vespignani.
\newblock {\em Dynamical processes on complex networks}.
\newblock Cambridge university press, 2008.

\bibitem{Kikuchi1966}
R.~Kikuchi.
\newblock The path probability method.
\newblock {\em Prog. Theor. Phys.}, 35:1, 1966.

\bibitem{MataFerreira2013}
A.~S. Mata and S.~C. Ferreira.
\newblock Pair quenched mean-field theory for the
  susceptible-infected-susceptible model on complex networks.
\newblock {\em Europhys. Lett.}, 103:48003, 2013.

\bibitem{Pelizzola2013}
A.~Pelizzola.
\newblock Variational approximations for stationary states of
  \textsc{I}sing-like models.
\newblock {\em Eur. Phys. J. B}, 86:120, 2013.

\bibitem{DominguezDelferraroRiccitersenghi2016}
E.~Dominguez, G.~Del~Ferraro, and F.~Ricci-Tersenghi.
\newblock A simple analytical description of the non-stationary dynamics in
  \protect{Ising} spin systems.
\newblock {\em J. Stat. Mech.: Theor. Exp.}, 2017:033303, 2017.

\bibitem{WadaKaburagi1994}
K.~Wada and M.~Kaburagi.
\newblock Relaxation kinetics of the path probability method.
\newblock {\em Prog. Theor. Phys.}, 115:273, 1994.

\bibitem{An1988}
G.~An.
\newblock A note on the cluster variation method.
\newblock {\em J. Stat. Phys.}, 52:727, 1988.

\bibitem{HeskesAlbersKappen2003}
T.~Heskes, K.~Albers, and B.~Kappen.
\newblock Approximate inference and constrained optimization.
\newblock In {\em Proceedings of the Nineteenth Conference on Uncertainty in
  Artificial Intelligence}, UAI'03, page 313, San Francisco, CA, USA, 2003.
  Morgan Kaufmann Publishers Inc.

\bibitem{PastorsatorrasVespignani2001}
R.~Pastor-Satorras and A.~Vespignani.
\newblock Epidemic spreading in scale-free networks.
\newblock {\em Phys. Rev. Lett.}, 86:3200, 2001.

\bibitem{MatsudaOgitaSasakiSato1992}
H.~Matsuda, N.~Ogita, A.~Sasaki, and K.~Sato.
\newblock Statistical mechanics of populations--\protect{The lattice
  Lotka-Volterra} model.
\newblock {\em Prog. Theor. Phys.}, 88:1035, 1992.

\bibitem{Moreno_et_al2010}
S.~G\'omez, A.~Arenas, J.~Borge-Holtoefer, S.~Meloni, and Y.~Moreno.
\newblock Discrete-time \protect{Markov} chain approach to contact-based
  disease spreading in complex networks.
\newblock {\em Europhys. Lett.}, 89:38009, 2010.

\bibitem{NeriBolle2009}
I.~Neri and D.~Boll\'{e}.
\newblock The cavity approach to parallel dynamics of \protect{Ising} spins on
  a graph.
\newblock {\em J. Stat. Mech.: Theor. Exp.}, 2009:P08009, 2009.

\bibitem{DelferraroAurell2015}
G.~Del~Ferraro and E.~Aurell.
\newblock Dynamic message-passing approach for kinetic spin models with
  reversible dynamics.
\newblock {\em Phys. Rev. E}, 92:010102, 2015.

\bibitem{Delferraro2016}
G.~Del~Ferraro.
\newblock {\em Equilibrium and dynamics on complex networks}.
\newblock PhD thesis, KTH Royal Institute of Technology, 2016.

\bibitem{AurellMahmoudi2011}
E.~Aurell and H.~Mahmoudi.
\newblock A message-passing scheme for non-equilibrium stationary states.
\newblock {\em J. Stat. Mech.: Theor. Exp.}, 2011:P04014, 2011.

\bibitem{AurellMahmoudi2011ctp}
E.~Aurell and H.~Mahmoudi.
\newblock Three lemmas on dynamic cavity method.
\newblock {\em Commun. Theor. Phys.}, 56:157, 2011.

\bibitem{AurellMahmoudi2012}
E.~Aurell and H.~Mahmoudi.
\newblock Dynamic mean-field and cavity methods for diluted \protect{Ising}
  systems.
\newblock {\em Phys. Rev. E}, 85:031119, 2012.

\bibitem{BarthelDebaccoFranz2015}
T.~Barthel, C.~De~Bacco, and S.~Franz.
\newblock A matrix-product algorithm for stochastic dynamics on locally
  tree-like graphs.
\newblock {\em ArXiv preprint arXiv:1508.03295}, 2015.

\bibitem{ShresthaScarpinoMoore2015}
M.~Shrestha, S.~V. Scarpino, and C.~Moore.
\newblock Message-passing approach for recurrent-state epidemic models on
  networks.
\newblock {\em Phys. Rev. E}, 92:022821, 2015.

\end{thebibliography}

\end{document}